\documentclass[english,sectionbib, sort&compress]{elsarticle}
\usepackage[T1]{fontenc}
\usepackage[latin9]{inputenc}
\usepackage{geometry}
\usepackage{array}
\usepackage{mathtools}
\usepackage{amsmath}
\usepackage{amssymb}
\usepackage{stackrel}
\usepackage{graphicx}

\makeatletter

\providecommand{\tabularnewline}{\\}

\usepackage[noline,noend,ruled]{algorithm2e}
\SetKwInput{KwInit}{Init}
\usepackage{algpseudocode}
\usepackage{multicol}
\usepackage{xspace}
\usepackage{pdfwidgets}
\usepackage{enumerate}
\newtheorem{thm}{Theorem}[section]
\newtheorem{lem}{Lemma}[section]
\newtheorem{cor}{Corollary}[section]
\newtheorem{rem}{Remark}[section]
\newdefinition{dfn}{Definition}[section]
\newproof{pf}{Proof}
\journal{Information Sciences}
\usepackage{hyperref}

\usepackage{etoolbox}
\patchcmd{\ps@pprintTitle}
{Preprint submitted to}{Accepted manuscript at}{}{}

\makeatother

\usepackage{babel}
\begin{document}

\begin{frontmatter}{}

\title{A theory of incremental compression}

\author{Arthur Franz\corref{cor1}}
\ead{af@occam.com.ua}
\author{Oleksandr Antonenko}
\ead{aa@occam.com.ua}
\author{Roman Soletskyi}
\ead{rs@occam.com.ua}
\cortext[cor1]{Corresponding author}
\address{Odessa Competence Center for Artificial intelligence and Machine learning (OCCAM), Odessa, Ukraine}
\tnotetext[t1]{\href{https://doi.org/10.1016/j.ins.2020.08.035}{https://doi.org/10.1016/j.ins.2020.08.035}  \newline \copyright \  2020. This manuscript version is made available under the CC-BY-NC-ND 4.0 license \href{http://creativecommons.org/licenses/by-nc-nd/4.0/}{http://creativecommons.org/licenses/by-nc-nd/4.0/}}
\begin{abstract}
The ability to find short representations, i.e.\ to compress data,
is crucial for many intelligent systems. We present a theory of incremental
compression showing that arbitrary data strings, that can be described
by a set of features, can be compressed by searching for those features
incrementally, which results in a partition of the information content
of the string into a complete set of pairwise independent pieces.
The description length of this partition turns out to be close to
optimal in terms of the Kolmogorov complexity of the string. Exploiting
this decomposition, we introduce ALICE -- a computable ALgorithm
for Incremental ComprEssion -- and derive an expression for its time
complexity. Finally, we show that our concept of a feature is closely
related to Martin-Löf randomness tests, thereby formalizing the meaning
of ``property'' for computable objects.
\end{abstract}

\end{frontmatter}{}

\section{Introduction}

In the machine learning community it has long been known that short
representations of data lead to high generalization abilities \citep{rissanen1978modeling,bishop2006pattern}.
If the model is larger than necessary the less it is able to generalize,
hence to predict.\footnote{Even though neural networks appear to grow in size every year, the
amount of data they are trained on is growing as well. Further, numerous
regularization techniques, weight sharing, sparse coding etc.\ are
being used in order to keep the networks' description length short.}

Similar ideas have been expressed in cognitive neuroscience, psychology
and linguistics. In neuroscience, the efficient coding hypothesis
\citep{barlow1961possible} states that the spikes in the sensory
system form a neural code minimizing the number of spikes needed to
transmit a given signal. Sparse coding \citep{olshausen1996emergence}
is the representation of items by the strong activation of a relatively
small set of neurons. In psychology, it is not a recent idea to regard
perception from the perspective of Bayesian inference \citep{knill1996perception},
which is known to entail Occam's razor \citep[Chapter 28]{mackay2003information}.\footnote{In a nutshell, the reason is the following. Any prior distribution
has to sum to $1$. For models with many parameters this probability
mass is smeared over a large space leading to a small prior probability
for any particular set of parameters, on average. See provided reference
for details.} In linguistics, various principles of information maximization and
communication cost minimization have been proposed (see \citep{kemp2012kinship}
for a review). In the philosophy of science, Occam's razor suggests
that scientists should strive for explanations / theories that are
as simple as possible while capturing and explaining as much data
as possible \citep{gauch2003scientific}.

Kolmogorov, Solomonoff and Chaitin went further and formalized those
ideas, which has led to the development of the algorithmic information
theory \citep{kolmogorov1965three,solomonoff1964formal1,solomonoff1964formal2}.
The amount of information contained in an object $x$ is defined as
the length $K(x)$ of its shortest possible description, which has
become known as the (prefix) Kolmogorov complexity. $K(x)$ is the
length of the shortest program that prints $x$ when executed on a
universal (prefix) Turing machine $U$, i.e.\ a computer. For example,
if $x$ is a string of one million zeros, we can write a short program
$q$ that can print it while being much shorter than the data, $l(q)\ll l(x)$.

Finding short descriptions for data is what data compression is all
about, since the original data can be unpacked again from its short
description. Furthermore, compression is closely tied to prediction.
For example, a short program implementing a zero printing loop could
just as well continue printing more than a million of them, which
would constitute a prediction of the continuation of the sequence.
Indeed, Solomonoff's theory of universal induction proves formally
that compressing data leads to the best possible predictor with respect
to various optimality criteria (see \citep[Chapter 3.6]{Hutter04uaibook})
in the set of lower semicomputable semimeasures \citep{solomonoff1978complexity}.
In order to do so, the so-called universal prior of a data string
$x$ is defined:

\begin{equation}
M(x)=\sum_{q:U(q)=x}2^{-l(q)}\label{eq:solprior}
\end{equation}
where $x,q\in\mathcal{B}^{*}$ are finite strings defined on a finite
alphabet $\mathcal{B}$, $U$ is a universal Turing machine that executes
program $q$ and prints $x$ and $l(q)$ is the length of program
$q$. Given already seen history $x_{<k}\equiv x_{1}\cdots x_{k-1}$
the predictor's task is to compute a probability distribution over
$x_{k}$, which is given by the conditional distribution $M\left(x_{k}\mid x_{<k}\right)=M\left(x_{\le k}\right)/M\left(x_{<k}\right)$.
The Solomonoff predictor has been shown to converge quickly \citep{solomonoff1978complexity}
to the true data generating distribution, allowing it to predict future
data with the least possible loss in the limit. Note that eq.\ (\ref{eq:solprior})
weighs each ``explanation'' $q$ for the data with $2^{-l(q)}$,
which directly expresses Occam's razor (i.e.\ compression): even
though we should consider all explanations, the shorter/simpler ones
should receive the highest weight. Remarkably, optimal induction and
prediction requires halving the prior probability of an explanation
for every additional bit in the explanation length.

In the context of artificial intelligence, Hutter went further and
attached the Solomonoff predictor to a reinforcement learning agent
\citep{Hutter04uaibook}. If general intelligence is defined as the
ability to achieve goals in a wide range of environments \citep{legg2007universal},
the resulting AIXI agent has been shown to exhibit maximal general
intelligence according to various optimality criteria \citep[Theorems 5.23 and 5.24]{Hutter04uaibook}.
This result formally ties the conceptual problem of artificial general
intelligence to efficient data compression, making the search for
its solution even more urgent.

Even though Kolmogorov complexity solves the task of optimal compression
from a theoretical point of view, its fundamental incomputability
as well as the impossibility of computing even approximate estimates
\citep[Theorem 2.3.2]{li2009introduction} hinders its practical usefulness.
Nevertheless, there is a modified computable measure -- Levin complexity
$Kt(x)$, which also takes into account the running time for a program
generating the string $x$. The well-known Levin Search \citep{levin1973universal}
algorithm computably finds a description with the lowest Levin complexity
and thereby constitutes in some sense a practical analogue of Kolmogorov
complexity. We should however note, that Levin complexity can in some
cases differ dramatically from the Kolmogorov complexity which is
an insurmountable consequence of the incomputability of the latter.
There is a number of algorithms trying to optimize Levin Search, like
Hutter Search, Adaptive Levin Search, the Optimal Ordered Problem
Solver and others \citep{Hutter:01fast,schmidhuber1997shifting,schmidhuber2004optimal}.
The problem is however, that the cited algorithms try to find the
entire description of the string $x$ at once, leading to very large
multiplicative or additive constants arising from the exhaustive search
through a large number of programs. This is naturally related to the
fact that in the definition of the Kolmogorov complexity $K(x)$ it
is required to find a single program describing the string. The search
for such a program can take a lot of time even if it runs quickly.
The main idea of the present paper is to express the complexity of
the string as a decomposition into several programs, which can be
obtained incrementally one after the other.

The main contributions of this paper are summarized as follows:
\begin{itemize}
\item The theory of (lossless) incremental compression is presented in full
scale here. We show that the information content (Kolmogorov complexity)
of an arbitrary incrementally compressible string -- the golden standard
of compression -- can be broken down into several mutually independent
pieces: the features of the string and a residual part. Some theorems
on this idea have already been published in a short conference paper
\citep{franz2016some}. In this paper, we extend those ideas substantially.\footnote{The conference paper unfortunately contains several mistakes even
though the core ideas are still valid. We have corrected them here.} We also treat the important question about the number of compression
steps (Chapter \ref{subsec:On-the-number}), discuss of the relationship
to machine learning and the application of the theory in practice.
\item The understanding of a feature is expanded by two degenerate notions:
the singleton and the universal feature. The former helps understand
that any compressible string possesses a feature -- the singleton
feature, making the existence of feature a ubiquitous property of
compressible strings. The latter leads to a constant bound on the
length of the shortest feature and descriptive map in the important
special case of well-compressible strings. This helps understand that
features can be very short, in particular much shorter than $K(x)$
which is crucial for the practical search for features. 
\item Our understanding of features is further deepened by Theorems \ref{thm:number_of_features}
and \ref{thm:independence-general}. Theorem \ref{thm:number_of_features}
expands our knowledge about the number of shortest features, which
turns out to grow at most polynomially with the length of the to-be-compressed
string. Theorem \ref{thm:independence-general} informs us about the
penalty to be paid (in bits of superfluous description length) if
the chosen feature is not the shortest, which is crucial if, for example,
incremental compression is implemented by a layered neural network
in which many bits are wasted in the weights between the layers.
\item The introduction of well-compressible strings leads to different compression
schemes discussed and compared in Chapter \ref{subsec:well-compressible},
leading to the introduction of ALICE in Chapter \ref{sec:Computable-IC}
-- a computable ALgorithm for Incremental ComprEssion. We derive
an expression for its time complexity.
\item Finally, we consolidate the notion of a feature by mapping it to Martin-Löf
randomness tests and at the same time establish a richer concept than
those tests, since features not only reflect the \textit{amount} of
randomness deficiency (as randomness tests do) but also the \textit{content}
of randomness deficiency. This finally justifies that features indeed
appear to formalize an algorithmic notion of a ``property'' of a
string.
\end{itemize}
We proceed as follows. In Chapter \ref{sec:Main-def-prop} we introduce
the main idea and examine the basic properties of the notion of a
feature of a data string. We explore the properties of a single compression
step in Chapter \ref{subsec:Properties-of-single-step}. Iterating
this step many times leads to the formulation of our compression scheme
in Chapters \ref{subsec:Incremental-compression-scheme}--\ref{subsec:Comparison-of-compression}.
We discuss computable incremental compression in Chapter \ref{sec:Computable-IC}.
Finally, we set up a bridge to Martin-Löf's theory of randomness in
Chapter \ref{subsec:Relationship-to-Martin-L=0000F6f}, grounding
the notion that a feature materializes a certain non-random property. 

\section{Main idea, definitions and basic properties\label{sec:Main-def-prop}}

\subsection{Main idea\label{subsec:Main-idea}}

In practice, describing an object often comes down to identifying
particular properties or features of the object, such as color, shape,
size, location etc. Intuitively, a description appears most accomplished
if those features are independent, i.e.\ do not contain information
about each other. For example, knowing the color of an object does
not contain information about its shape and vice versa. This independence
appears to allow us to find the features one by one, incrementally,
without having to find the full description at once.

\begin{figure}[h]
\centering{}\includegraphics{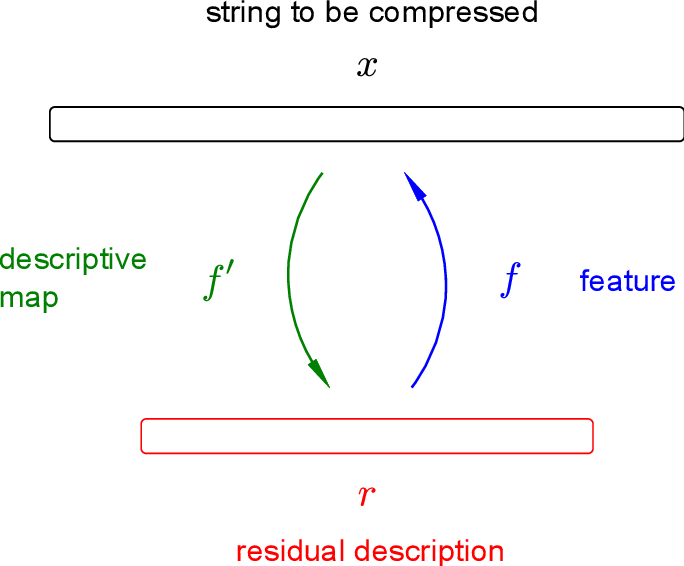}\caption{\label{fig:autoencoder}An autoencoder.}
\end{figure}

In order to formalize this idea, we represent any data string $x$
by a composition of functions, $x=\left(f_{1}\circ\cdots\circ f_{s}\right)(r_{s})$,
by looking for stacked autoencoders \citep{hinton2006reducing}. The
idea of an autoencoder is to use a \textit{descriptive map} $f'$
to project input data $x$ on a shorter \textit{residual description
}$r$, from which a \textit{feature} map $f$ can reconstruct it,
see Fig.\ \ref{fig:autoencoder}. The idea was inspired by SS'-Search
\citep{potapovSSsearch}.

\paragraph{Examples}

Consider a string of the form $x=1^{n}0y$ (where $1^{n}:=\overbrace{1\ldots1}^{n\text{ times}}$).
Then, a descriptive map could be a function computing the number $n$
of initial ones and copying $y$ to the residual description, $f_{1}'(x)=\left\langle n,y\right\rangle =:r$.
A feature could be a function taking the residual $r=\left\langle n,y\right\rangle $
which is a combination of number of ones $n$ and the rest of string
$y$ and mapping it back to $x$, $f_{1}(r)=x$. Clearly, the residual
description is much shorter than $x$ for large enough $n$, meaning
that some data compression has been achieved (encoding a number $n$
as a binary string takes only logarithmic size of $n$ which is much
less than $n$ bits).

This example demonstrates the existence of some fixed feature of an
infinite number of binary strings. If the string $y$ has a similar
form as $x$, say $y=1^{m}0z$, then for a sufficiently large $m$
the function $f_{2}\left(\left\langle n,m,z\right\rangle \right)=\left\langle n,y\right\rangle $
will be a feature of the residual $r=\left\langle n,y\right\rangle $,
where $\left\langle n,m,z\right\rangle $ denotes some fixed encoding
of the triple $n,m,z$. In this way, $x=1^{n}01^{m}0z=f_{1}\left(f_{2}\left(\left\langle n,m,z\right\rangle \right)\right)$
where the residual description $\left\langle n,m,z\right\rangle $
is much shorter than $x$.

Consider a string of the form $w=1^{n}01^{n+1}01^{n+2}0\ldots1^{n+m}0$
for sufficiently large $n$ and $m$. On the one hand, the above reasoning
can be continued leading to the composition $w=\left(f_{1}\circ\cdots\circ f_{m+1}\right)\left(\left\langle n,n+1,\ldots,n+m,\epsilon\right\rangle \right)$,
where $f_{i}\left(\left\langle a_{1},a_{2},\ldots,a_{i-1},a_{i},z\right\rangle \right)=\left\langle a_{1},a_{2},\ldots,a_{i-1},1^{a_{i}}0z\right\rangle $
and $\epsilon$ is the empty string. On the other hand, we can consider
a combined feature like $f\left(\left\langle k,\left\langle a_{1},a_{2},\ldots,a_{k}\right\rangle \right\rangle \right)=1^{a_{1}}01^{a_{2}}0\ldots1^{a_{k}}0$.
The corresponding descriptive map $f'\left(w\right)$ counts the numbers
$a_{1},a_{2},\ldots,a_{k}$ of consecutive ones in $w$ as well as
the number of such groups $k$. The residual description $\left\langle k,\left\langle a_{1},a_{2},\ldots,a_{k}\right\rangle \right\rangle $
will be substantially shorter than $w$ for large $a_{1},a_{2},\ldots,a_{k}$,
i.e.\ when the number of ones is much larger than the number of zeros.
Then, $r=f'(w)=f'\left(1^{n}01^{n+1}01^{n+2}0\ldots1^{n+m}0\right)=\left\langle m+1,\left\langle n,n+1,\ldots,n+m\right\rangle \right\rangle $.
Moreover, defining a feature according to $g\left(\left\langle n,m\right\rangle \right)=\left\langle m+1,\left\langle n,n+1,\ldots,n+m\right\rangle \right\rangle $
leads to the concise representation $w=f\left(g\left(\left\langle n,m\right\rangle \right)\right)$.

\begin{figure}[h]
\centering{}\includegraphics{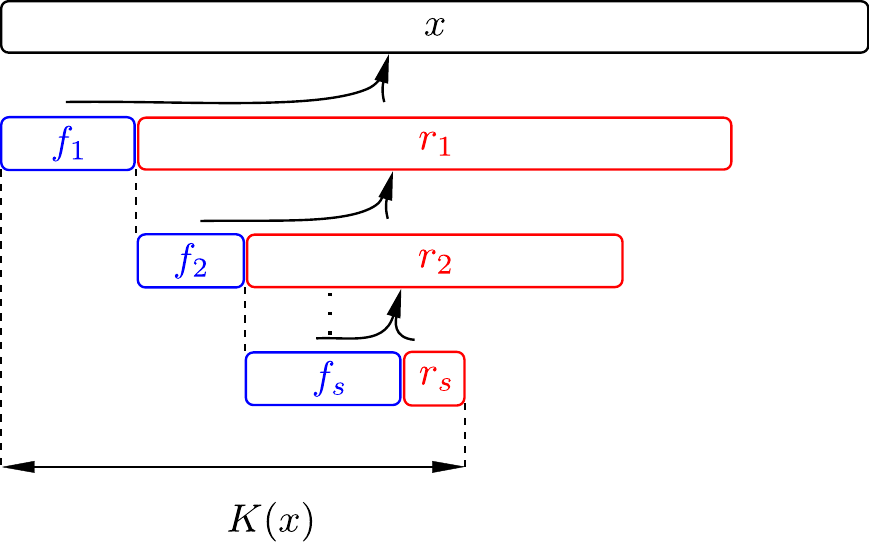}\caption{\label{fig:bounds_on_complexity}A string $x$ is incrementally compressed,
i.e.\ its description length decreases every step until the shortest
description $K(x)$ is approximated by $\sum_{i=1}^{s}l\left(f_{i}\right)+K\left(r_{s}\right)$,
see Theorem \ref{thm:compression_scheme} below. The compression condition
$l\left(f_{i}\right)+l\left(r_{i}\right)<l\left(r_{i-1}\right)$ ensures
that the description length of each residual $r_{i}$ decreases.}
\end{figure}

The key advantage of using an autoencoder is that a shorter representation
$r$ of the data $x$ can be arbitrarily computed from $x$ instead
of restricting the forms of computation to exhaustive search for $r$.
This allows for the possibility that $r$ can be found much faster
than by search. In the just discussed example, the residual is found
by counting the numbers of ones in the string instead of trying to
guess them. At the same time, information in $r$ about $x$ is preserved
by requiring the ability to reconstruct it back from $r$. Instead
of searching for a potentially long description $r$, the search is
focused on finding a pair of hopefully simple functions $(f,f')$
that manage to compress $x$ at least a little bit, which we call
the compression condition. A little compression shall be enough, since
this process is repeated for the residual description $r$ in the
role of input data for the next autoencoder required to compress $r$
a little further. The process iterates until step $s$ where no compression
is possible, as depicted in Fig.\ \ref{fig:bounds_on_complexity}.

\subsection{Preliminaries}

Consider strings made up of elements of the set $\mathcal{B}=\left\{ 0,1\right\} $
with $\epsilon$ denoting the empty string. $\mathcal{B}^{\ast}$
denotes the set of finite strings over $\mathcal{B}$. Denote the
length of a string $x$ by $l(x)$. Since there is a bijective map
$\mathcal{B}\rightarrow\mathcal{N}$ of finite strings onto natural
numbers, strings and natural numbers are used interchangeably. We
define the prefix-codes $E_{1}\left(x\right)=\overline{x}=1^{l(x)}0x$
and $E_{2}(x)=\overline{l(x)}x$.

The universal, prefix Turing machine $U$ is defined by 
\begin{equation}
U\left(\left\langle y,\left\langle i,q\right\rangle \right\rangle \right)=T_{i}\left(\left\langle y,q\right\rangle \right),
\end{equation}
where $T_{i}$ is $i$-th machine in some enumeration of all prefix
Turing machines and $\left\langle .,.\right\rangle $ is an one-to-one
mapping $\mathcal{N\times\mathcal{N}\rightarrow\mathcal{N}}$, defined
by $\left\langle x,y\right\rangle \equiv\overline{x}y$. This means
that $f=\left\langle i,q\right\rangle $ describes some program on
a prefix Turing machine and consists of the number $i$ of the machine,
its input $q$ and some additional parameter $y$. We will use the
shortcut $f(y):=U\left(\left\langle y,f\right\rangle \right)$ defining
how strings, interpreted as Turing machines (partial recursive functions),
can execute other strings. The conditional (prefix Kolmogorov) complexity
is given by
\begin{equation}
K(x\mid r):=\mbox{min}_{f}\left\{ l(f):\;U\left(\left\langle r,f\right\rangle \right)\equiv f(r)=x\right\} 
\end{equation}
This means that $K(x\mid r)$ is the length of the shortest function
$f$ able to compute $x$ from $r$ on a universal prefix Turing machine.
The unconditional (prefix Kolmogorov) complexity is defined by $K(x)\equiv K(x\mid\epsilon)$.
We define the information in $x$ about $y$ as $I(x:y):=K(y)-K(y\mid x)$.
Up to this point, we have followed the standard definitions as given
in \citep{li2009introduction}.

\subsection{Definitions}

\begin{dfn}\label{definition1}Let $f$ and $x$ be finite strings
and $D_{f}\left(x\right)$ the set of \textbf{descriptive maps} of
$x$ given $f$:
\begin{equation}
D_{f}\left(x\right)=\left\{ f^{\prime}:\,f\left(f^{\prime}\left(x\right)\right)=x,\,l\left(f\right)+l\left(f^{\prime}\left(x\right)\right)<l\left(x\right)\right\} \label{eq:set-of-desc-maps}
\end{equation}
If $D_{f}\left(x\right)\neq\emptyset$ then $f$ is called a \textbf{feature}
of $x$. The strings $r\equiv f'(x)$ are called \textbf{residual
descriptions} of the feature $f$. $f^{*}$ is called \textbf{shortest
feature} of $x$ if it is one of the strings fulfilling
\begin{equation}
l\left(f^{*}\right)=\min\left\{ l(f):\;D_{f}(x)\neq\varnothing\right\} 
\end{equation}
and $f'^{*}$ is called \textbf{shortest descriptive map} of $x$
given $f^{*}$ if
\begin{equation}
l\left(f'^{*}\right)=\min\left\{ l(g):\;g\in D_{f^{*}}(x)\right\} \label{eq:shdesc-1}
\end{equation}
The so-called \textbf{compression condition} 
\begin{equation}
l\left(f\right)+l\left(r\right)<l\left(x\right)\label{eq:Compresion_Condition}
\end{equation}
is required to avoid the case of $f$ and $f'$ being the identity
functions in eq.\ (\ref{eq:set-of-desc-maps}), leading to the useless
transformation $r=id(x)=x$. \end{dfn}

Note that in eq.\ (\ref{eq:Compresion_Condition}) $f$ has to be
part of the description of $x$, since otherwise $r$ can be made
arbitrarily small by inserting all the information into $f$. 

\subsection{Basic properties}

Our first result shows that features of compressible strings always
exist, since the shortest description of $x$ could be coded into
a single function $f$ using an empty residual $r=\epsilon$. We shall
call this a \textbf{singleton feature} of the string:

\begin{lem}\label{lem:lemma_total_feature}If $x$ is compressible,
that is $K\left(x\right)<l\left(x\right)$, then there exists a singleton
feature $f$, and its residual $r=\epsilon$ such that $f(r)=x$,
$l\left(f\right)+l\left(r\right)=K\left(x\right)<l\left(x\right)$.
The length of the shortest feature is $l\left(f^{\ast}\right)\leq K\left(x\right)$.\end{lem}

\begin{pf}Let $K\left(x\right)=K\left(x\mid\epsilon\right)<l\left(x\right)$,
then there exists a string $f$ such that $U\left(\left\langle \epsilon,f\right\rangle \right)=f\left(\epsilon\right)=x$,
$l\left(f\right)=K\left(x\right)$. Clearly, $f$ is a feature, since
with $r=\epsilon$ the compression condition $l\left(f\right)+l\left(r\right)=K\left(x\right)+0<l\left(x\right)$
is fulfilled. Thus for the shortest feature $f^{\ast}$ we obtain
$l\left(f^{\ast}\right)\leq l\left(f\right)=K\left(x\right)$.\qed\end{pf}

On the other extreme, the universal Turing machine $U$ itself could
function as a feature function, since nothing prevents a universal
machine to simulate another universal machine. Such features shall
be called universal, since they are able to compute any computable
$x$. However, $x$ has to be compressible enough to accommodate the
length of $U$ in order to fulfill the compression condition.

\begin{dfn}\label{def:universal-feature}A string $f_{0}$ shall
be called \textbf{universal feature}, if there is a constant $C$,
such that $f_{0}$ is a feature of any string compressible by more
than C bits.\end{dfn}

\begin{lem}\label{lem:lemma_universal_feature} There exists a constant
$C$ and a universal feature $f_{0}$ with length $l\left(f_{0}\right)=C$
such that $f_{0}$ is a feature of any string $x$ compressible by
more than $C$ bits: $K\left(x\right)<l\left(x\right)-C$. The residual
description $r$ is a shortest description of $x$, $l\left(r\right)=K\left(x\right)$.\end{lem}

\begin{pf}Consider some finite string $x$ and its shortest description
$\left\langle i,t\right\rangle $. Then $U\left(\left\langle \epsilon,\left\langle i,t\right\rangle \right\rangle \right)=T_{i}\left(\left\langle \epsilon,t\right\rangle \right)=x$
and $l\left(\left\langle i,t\right\rangle \right)=K\left(x\right)$.
Define $h\left(\left\langle \left\langle i,t\right\rangle ,\epsilon\right\rangle \right)=T_{i}\left(\left\langle \epsilon,t\right\rangle \right)$,
which is a kind of universal Turing machine. Let its number be $j$
in the enumeration of Turing machines, $h=T_{j}$. We obtain 
\begin{equation}
U\left(\left\langle \left\langle i,t\right\rangle ,\left\langle j,\epsilon\right\rangle \right\rangle \right)=T_{j}\left(\left\langle \left\langle i,t\right\rangle ,\epsilon\right\rangle \right)=T_{i}\left(\left\langle \epsilon,t\right\rangle \right)=x.
\end{equation}
Denote $f_{0}=\left\langle j,\epsilon\right\rangle $, $r=\left\langle i,t\right\rangle $,
$C=l\left(f_{0}\right)$. Then, $f_{0}\left(r\right)=x$, $l\left(r\right)=K\left(x\right)$,
and $l\left(f_{0}\right)+l\left(r\right)=C+K\left(x\right)<l\left(x\right)$,
since $x$ is compressible by more than $C$ bits by assumption. Therefore,
$f_{0}$ is a feature of $x$. \qed\end{pf}

\begin{rem}\label{rem:remark1} Lemma \ref{lem:lemma_universal_feature}
shows that the length $l\left(f^{\ast}\right)$ of the shortest feature
is not just bounded by the complexity $K\left(x\right)$ but can be
substantially shorter. In the case of strings satisfying Lemma \ref{lem:lemma_universal_feature}
it is bounded by a constant $l\left(f_{0}\right)=C$ whereas the complexity
$K(x)$ can be arbitrary large. Note that $l\left(f_{0}\right)=C$
depends on the choice of the universal Turing machine $U$. If we
choose such a machine that makes that constant small, by taking $j=0$
for example, then for a large number of strings $x$ compressible
by more than $C$ bits the universal feature $f_{0}$ will be the
shortest one. Nevertheless, by some natural choice of $U$ we can
expect that the universal feature $f_{0}$ will be sufficiently long
to allow features shorter than $l\left(f_{0}\right)$ for many strings.\end{rem}

\section{Incremental compression}

\subsection{Properties of a single compression step\label{subsec:Properties-of-single-step}}

\begin{figure}
\centering{}\includegraphics[width=1\textwidth]{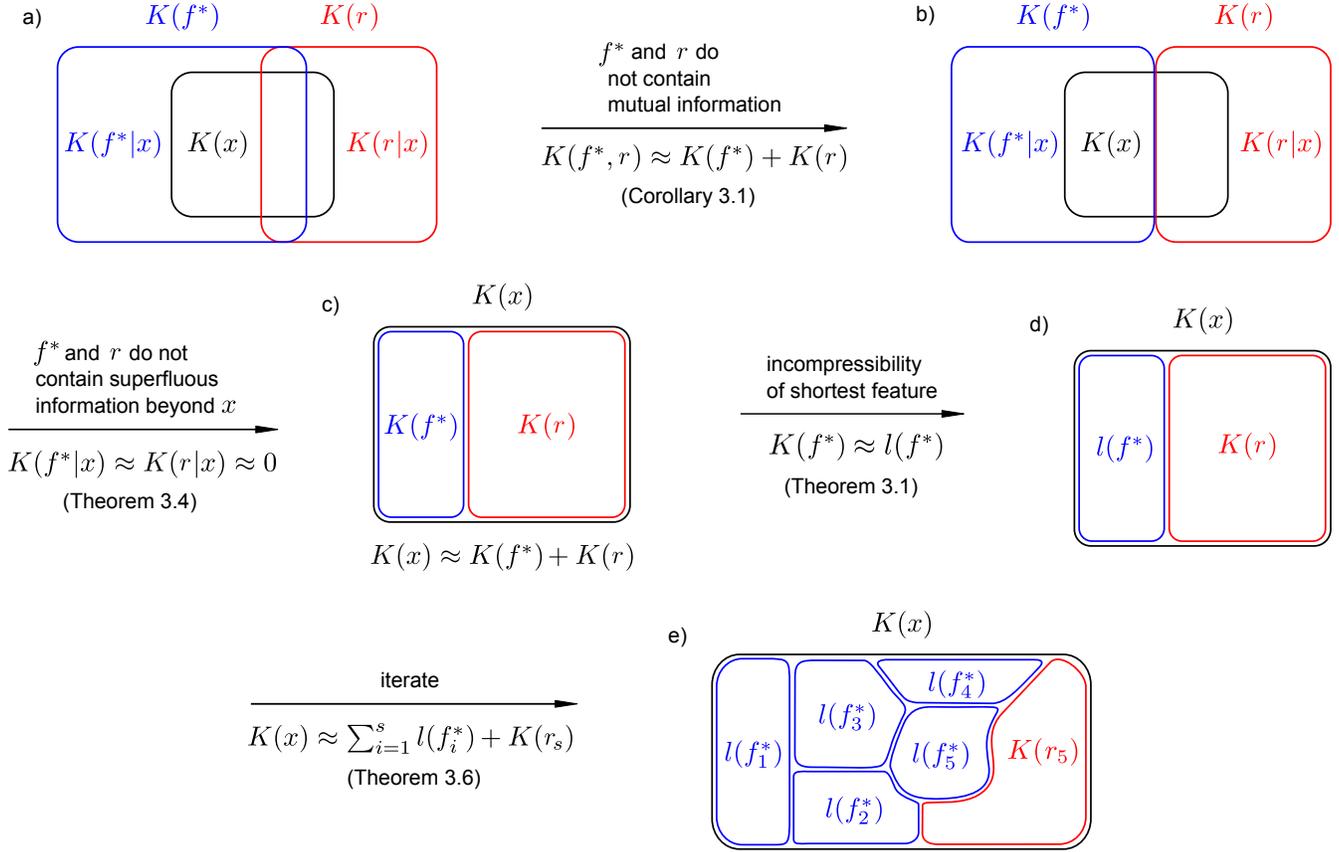}\caption{\label{fig:theorems} The proof strategy. a) Since $x$ is computed
from shortest feature $f^{*}$ and residual $r$, $f^{*}(r)=x$, all
its information is contained in the union of $f^{*}$ and $r$. b)
Theorem \ref{thm:independence-general} and Corollary \ref{cor:independence}
show that $f^{*}$ and $r$ do not contain mutual information and
therefore do not overlap. c) Theorem \ref{thm:no_superfluous_information}
shows that $f^{\ast}$ and $r$ do not contain information beyond
what is necessary to compute $x$. It follows that the information
in $x$ is partitioned into the information in $f^{\ast}$ and in
$r$. d) Theorem \ref{thm:feature_incompressibility} shows that the
shortest feature is incompressible, hence its length $l(f^{\ast})$
coincides with the amount of information it contains. e) Finally,
repeating the process on $r$ partitions the information inside $x$
into many mutually independent and incompressible pieces (Theorem
\ref{thm:compression_scheme}).}
\end{figure}

In the following, we will show that the information $K(x)$ inside
$x$ is partitioned by the shortest features $f_{1}^{*},\ldots f_{s}^{*}$
and the last residual $r_{s}$ by the process of incremental compression.
The main proof strategy is illustrated in Fig.\ \ref{fig:theorems}.

What is the consequence of picking the shortest possible feature?
It turns out that similarly to shortest descriptions per se, the shortest
features are themselves incompressible. After all, if the shortest
feature was compressible, there would be a shorter program $q$ to
compute it. But then, $q$ can just as well take $r$ and go on computing
$x$ while becoming a feature itself, shorter than the shortest feature,
$l(q)<l(f^{*})$, which is a contradiction. Theorem \ref{thm:feature_incompressibility}
formalizes this reasoning. In order to prove it, we will need the
following lemma:

\begin{lem}\label{lem:lemma_feature_length_complexity}Let $f^{\ast}$
and $f^{\prime\ast}$ be the shortest feature and shortest descriptive
map of a finite string $x$, respectively. Further, let $r\equiv f^{\prime\ast}\left(x\right)$.
Then $l\left(f^{\ast}\right)=K\left(x\mid r\right)$ and $l\left(f^{\prime\ast}\right)=K\left(r\mid x\right)$.\end{lem}

\begin{pf}

First, we prove $l(f^{*})=K(x\mid r)$. Suppose the opposite $K\ensuremath{\left(x\mid r\right)=\min\left\{ l\left(z\right):\,z\left(r\right)=x\right\} \neq l\left(f^{\ast}\right)}$.
This means that there exists a shorter program $g$ with $l\left(g\right)<l\left(f^{\ast}\right)$
such that $g\left(r\right)=f^{\ast}\left(r\right)=x$. We have to
check whether $D_{g}\left(x\right)\neq\emptyset$ holds. First, we
take a descriptive map $g^{\prime}:=f^{\prime\ast}$ and prove that
$g^{\prime}\in D_{g}\left(x\right)$: We have $g\left(g^{\prime}\left(x\right)\right)=g\left(f^{\prime\ast}\left(x\right)\right)=g\left(r\right)=x$
and $l\left(x\right)>l\left(f^{*}\right)+l\left(f^{\prime\ast}\left(x\right)\right)>l\left(g\right)+l\left(g^{\prime}\left(x\right)\right)$.
Therefore, $g^{\prime}\in D_{g}\left(x\right)$ and $D_{g}\left(x\right)\neq\emptyset$
implying that $g$ is a feature of $x$. However, since $f^{*}$ is
defined as the shortest feature, the assumption $l\left(g\right)<l\left(f^{\ast}\right)$
constitutes a contradiction.

Second, we prove $l\left(f^{\prime\ast}\right)=K(r\mid x)$. Suppose
the opposite $K\ensuremath{\left(r\mid x\right)=\min\left\{ l\left(z\right):\,z\left(x\right)=r\right\} \neq l\left(f^{\prime\ast}\right)}$.
This means that there exists a shorter program $g^{\prime}$ with
$l\left(g^{\prime}\right)<l\left(f^{\prime\ast}\right)$ such that
$g^{\prime}\left(x\right)=f^{\prime\ast}\left(x\right)=r$. Then $g^{\prime}\in D_{f^{\ast}}\left(x\right)$
since $f^{\ast}\left(g^{\prime}\left(x\right)\right)=f^{\ast}\left(r\right)=x$
and $l\left(f^{\ast}\right)+l\left(g^{\prime}\left(x\right)\right)=l\left(f^{\ast}\right)+l\left(f^{\prime\ast}\left(x\right)\right)<l\left(x\right)$,
which contradicts the assumption that $f^{\prime\ast}$ is the shortest
program in $D_{f^{\ast}}\left(x\right)$. \qed

\end{pf}

\begin{thm}[Feature incompressibility]\label{thm:feature_incompressibility}The
shortest feature $f^{\ast}$ of a finite string $x$ is incompressible:
\begin{equation}
l\left(f^{\ast}\right)-O\left(1\right)\leq K\left(f^{\ast}\right)\leq l\left(f^{\ast}\right)+K\left(l\left(f^{\ast}\right)\right)+O\left(1\right)\leq l\left(f^{\ast}\right)+O\left(\log l(f^{\ast})\right).
\end{equation}
\end{thm}\begin{pf}Since $x$ can be obtained from $f^{*}$ and
$r$, we get
\begin{equation}
l(f^{*})=K(x\mid r)\le K(f^{\ast}\mid r)+O\left(1\right)\le K(f^{\ast})+O\left(1\right)
\end{equation}

Moreover, the general property of prefix Kolmogorov complexity, $K(f^{\ast})\le l(f^{\ast})+K\left(l\left(f^{\ast}\right)\right)+O\left(1\right)\le l(f^{\ast})+O\left(\log l(f^{\ast})\right)$,
holds. \qed\end{pf}

The next theorem shows that the number of shortest features is actually
quite small. Intuitively, this is due to the fact that the compression
condition requires the features to do some actual compression work,
which is a rare capacity.

\begin{lem}\label{lem:K_f_x} Let $f$ be a feature of a finite string
$x$ and $r$ its residual description. Then,

\begin{equation}
K(x\mid f)\leq l(r)+O\left(\log l(r)\right)<l(x)-l(f)+O\left(\log l(x)\right).
\end{equation}

\end{lem}\begin{pf}If $f$ is a feature then by definition there
exists a residual $r$ such that $f(r)=x$ and $l(f)+l(r)<l(x)$.
Consider $K(x\mid f)$ which is somewhat analogous to $K(x\mid r)$
in Lemma \ref{lem:lemma_feature_length_complexity}. We can exchange
$f$ and $r$ by running $U\left(\left\langle f,\left\langle i,E_{2}(r)\right\rangle \right\rangle \right)=T_{i}\left(\left\langle f,E_{2}(r)\right\rangle \right)=U\left(\left\langle r,f\right\rangle \right)=x$,
where $T_{i}$ swaps $f$ and $r$. In order to be able to do so,
$r$ has to be encoded in a self-delimiting way with $E_{2}(r)$ for
example. Using the invariance theorem, we find that $K(x\mid f)\leq l\left(E_{2}(r)\right)+O(1)=l(r)+2l\left(l(r)\right)+O(1)$
leading to 
\begin{equation}
K(x\mid f)\leq l(r)+O\left(\log l(r)\right)<l(x)-l(f)+O\left(\log l(x)\right).
\end{equation}
\qed\end{pf}

\begin{lem}\label{lem:I_f_x} Let $f$ be a feature of a finite string
$x$ and $r$ its residual description. If $I(f:x)\coloneqq K(x)-K(x\mid f)\geq l(f)-O\left(\log l(x)\right)$
then

\begin{equation}
K(f\mid x)\leq O\left(\log l(x)\right).
\end{equation}

\end{lem}\begin{pf}Using the symmetry of algorithmic information
\citep[Lemma 3.9.2]{li2009introduction}
\begin{equation}
\left\lvert I(f:x)-I(x:f)\right\rvert \leq\log K(f)+\log K(x)+2\log\log K(f)+2\log\log K(x)+O(1)
\end{equation}
 and inequalities $K(x)\leq O(l(x))$ and $K(f)\leq O(l(f))\leq O(l(x))$
we obtain 
\begin{equation}
\left\lvert I(f:x)-I(x:f)\right\rvert \leq O\left(\log l(x)\right).
\end{equation}
It follows that $K(f)-K(f\mid x)=I(x:f)\geq I(f:x)-O\left(\log l(x)\right)\geq l(f)-O\left(\log l(x)\right)$.
Therefore, $K(f\mid x)\leq K(f)-l(f)+O\left(\log l(x)\right)\leq O\left(\log l(f)\right)+O\left(\log l(x)\right)\leq O\left(\log l(x)\right)$,
since $K(f)\leq l(f)+O\left(\log l(f)\right)$ and $l(f)<l(x)$.\qed\end{pf}

\begin{thm}[The number of shortest features]\label{thm:number_of_features}The
number of shortest features of a binary string $x$ grows at most
polynomially with $l(x)$. Moreover, $K(f^{*}\mid x)\leq O\left(\log l(x)\right)$
for any shortest feature $f^{*}$ of $x$.\end{thm}

\begin{pf}If $f$ is a feature then by Lemma \ref{lem:K_f_x} we
have $K(x\mid f)\leq l(x)-l(f)+O\left(\log l(x)\right)$. Let us consider
two cases: if $l(x)-K(x)>C$ for some constant $C=l\left(f_{0}\right)$
then a universal feature $f_{0}$ of string $x$ appears (see Definition
\ref{def:universal-feature} and Lemma \ref{lem:lemma_universal_feature}),
which limits number of \textit{shortest} features by some constant
(at most $2^{C+1}-1$) since their length is not greater than $l\left(f_{0}\right)=C$.
In this case, $K(f^{*}\mid x)\leq K(f^{*})+O\left(1\right)\leq l(f^{*})+2\log l(f^{*})+O(1)\leq C+2\log C+O(1)=O(1)$.

Otherwise, if $l(x)-K(x)\le C$ for some constant $C$, then $K(x\mid f)\leq l(x)-l(f)+O\left(\log l(x)\right)\leq K(x)+C-l(f)+O\left(\log l(x)\right)=K(x)-l(f)+O\left(\log l(x)\right)$
since the constant $C$ can be subsumed into $O\left(\log l(x)\right)$.
Thus, $I(f:x)\coloneqq K(x)-K(x\mid f)\geq l(f)-O\left(\log l(x)\right)$
and by Lemma \ref{lem:I_f_x} $K(f\mid x)\leq O\left(\log l(x)\right)$.
We know that the number of strings $z$ fulfilling $K(z\mid x)\leq a$
can not exceed the number of descriptions of maximal length $a$,
which is $2^{a+1}-1$. Therefore, for number of all features is not
greater than $2^{O\left(\log l(x)\right)+1}=2^{O\left(\log l(x)\right)}$,
which implies at most polynomial growth of the number of \emph{all}
features $O\left(l(x)^{\alpha}\right)$ for some constant $\alpha$.\qed\end{pf}

Now we move to a central result that a shortest feature does not contain
information about the residual and vice versa. This is important,
since otherwise, there would be an informational overlap between them,
indicating that we would be wasting description length on the overlap.
Figuratively speaking, if some ``bits'' inside $f^{*}$ describe
the same information as some ``bits'' in $r$, we could choose $f^{*}$
to be even shorter and avoid this redundancy. Interestingly, the number
of wasted bits is simply bounded by the difference in lengths, $l(f)-l(f^{*})$,
between the chosen feature and the shortest one:

\begin{thm}[Independence of features and residuals, general case]\label{thm:independence-general}Let
$f$ be a feature of a finite string $x$ and $r$ its residual description.
Further, let $f^{*}$ be a shortest feature and $d:=l(f)-l(f^{*})$.
Then,
\begin{align}
I(r:f)\coloneqq K(f)-K(f\mid r) & \leq d+K\left(l\left(f\right)\right)+O(1)\leq d+O(\log l(f)),\label{eq:I_p_f}\\
K\left(r\right)+K\left(f\right)-K\left(f,r\right) & \leq d+K\left(l\left(f\right)\right)+K\left(K\left(r\right)\mid r\right)+O\left(1\right)\nonumber \\
 & \leq d+O(\log l(f))+O(\log l(r))\label{eq:I_p_f2}
\end{align}
\end{thm}\begin{pf}

First, we prove eq.\ (\ref{eq:I_p_f}). We know that $f\left(r\right)=x$.
Consider the shortest string $\hat{f}$ with $\hat{f}\left(r\right)=x$
and $l\left(\hat{f}\right)=K(x\mid r)$. Then, $l\left(\hat{f}\right)\leq l\left(f\right)$
and the compression condition $l\left(\hat{f}\right)+l\left(r\right)\leq l\left(f\right)+l\left(r\right)\leq l\left(x\right)$
holds, since $f$ is a feature. Therefore, $\hat{f}$ is also a feature.
But since $f^{*}$ is a shortest feature we know $l(f^{*})\leq l\left(\hat{f}\right)\leq l\left(f\right)$
and $l(f)-l(\hat{f})\leq l(f)-l(f^{*})=:d$. Further, observe that
\begin{align}
K(f)\le l(f)+K\left(l\left(f\right)\right)+O(1)\le & d+l(\hat{f})+K\left(l\left(f\right)\right)+O(1)=\\
d+K(x\mid r)+K\left(l\left(f\right)\right)+O(1)\le & d+K(f\mid r)+K\left(l\left(f\right)\right)+O(1)
\end{align}
Here, the first and last inequalities are direct consequences of general
properties of the prefix Kolmogorov complexity and of the equation
$f\left(r\right)=x$. Thus, $I(r:f)\coloneqq K(f)-K(f\mid r)\leq d+K\left(l\left(f\right)\right)+O(1)\leq d+O(\log l(f))$.

Second, we prove eq.\ (\ref{eq:I_p_f2}). Using the general expansion
$K\left(f\mid r\right)\leq K\left(f\mid r,K\left(r\right)\right)+K\left(K\left(r\right)\mid r\right)+O\left(1\right)$
and (\ref{eq:I_p_f}) we have
\begin{align}
K(f,r)=K(r)+K\left(f\mid r,K\left(r\right)\right)+O\left(1\right)\ge & K(r)+K\left(f\mid r\right)-K\left(K\left(r\right)\mid r\right)-O\left(1\right)\ge\\
K\left(r\right)+K\left(f\right)-d-K\left(l\left(f\right)\right)-K\left(K\left(r\right)\mid r\right)-O\left(1\right)\geq & K\left(r\right)+K\left(f\right)-d-O(\log l(f))-O(\log l(r))
\end{align}
\qed\end{pf}

This theorem is valid for features in general. In particular, we obtain
for \textit{shortest} features the following relationships:

\begin{cor}[Independence of features and residuals]\label{cor:independence}Let
$f^{*}$ be the shortest feature a finite string $x$ and $r$ its
residual description. Then,
\begin{align}
K(f^{*}\mid r) & =K(f^{*})+O(\log l(f^{*})),\label{eq:fgr}\\
K(r\mid f^{*}) & =K(r)+O(\log l(f^{*}))+O(\log l(r)),\label{eq:rgf}\\
K(f^{*},r) & =K(f^{*})+K(r)+O(\log l(f^{*}))+O(\log l(r))\label{eq:f_ind_r}
\end{align}
\end{cor}\begin{pf}First, we prove eq.\ (\ref{eq:fgr}). We have
$K(f^{*})\leqslant K(f^{*}\mid r)+O(\log l(f^{*}))$$\leqslant K(f^{*})+O(\log l(f^{*}))$.
The first inequality follows from Theorem \ref{thm:independence-general}
in the case $f=f^{*}$, thus $d=0$. The second inequality is a general
property of Kolmogorov complexity.

As for eq.\ (\ref{eq:rgf}), consider the Kolmogorov-Levin theorem
to logarithmic precision (see, e.g.\ \citep[Theorem 21]{shen2017kolmogorov}
and note that the prefix and plain complexities coincide at logarithmic
precision):

\begin{equation}
K(f^{*},r)=K(f^{*})+K\left(r\mid f^{*}\right)+O(\log l(f^{*}))=K(r)+K\left(f^{*}\mid r\right)+O(\log l(r))\label{eq:fp_expansion-1}
\end{equation}
Using the just proven equality we obtain: $K(r\mid f^{*})=K(r)+K(f^{*}\mid r)-K(f^{*})+O(\log l(r))+O(\log l(f^{*}))=K(r)+O(\log l(r))+O(\log l(f^{*}))$

Finally, we turn to eq.\ (\ref{eq:f_ind_r}) and insert the just
obtained result into eq.\ (\ref{eq:fp_expansion-1}): 
\[
K(r,f^{*})=K(f^{*})+K(r\mid f^{*})+O(\log l(f^{*}))=K(f^{*})+K(r)+O(\log l(r))+O(\log l(f^{*}))
\]
\qed\end{pf}

We conclude that features and residuals do not share information about
each other, therefore the description of the $(f^{*},r)$-pair breaks
down into the simpler task of describing $f^{*}$ and $r$ separately.
Since Theorem \ref{thm:feature_incompressibility} implies the incompressibility
of $f^{*}$ and $U\left(\left\langle r,f^{*}\right\rangle \right)=x$,
the task of compressing $x$ is reduced to the mere compression of
$r$. Hence, we can sloppily write $K(f^{*},r)\approx l(f^{*})+K(r)$,
where the ``$\approx$'' sign denotes equality up to additive logarithmic
terms.

However, there is still a possibility that the $(f^{*},r)$-pair contains
\textit{more} information than necessary to compute $x$. As we shall
see this is not the case: no description length is wasted. Intuitively,
since the shortest feature $f^{*}$ neither overlaps with $r$, nor
contains redundant bits due to its incompressibility, the only way
for $f^{*}$ to contain superfluous information is to contain totally
unrelated noise with respect to $r$ and $x$. But that also does
not make sense if the shortest feature is picked. 

Apart from that we also show that the shortest descriptive map $l\left(f^{\prime*}\right)$
is very short. This is due to the fact that given $l\left(f^{*}\right)$
we can pick an algorithm $f^{\prime}$ of constant size that loops
through all possible pairs $(f,r)$ with $l(f)=l\left(f^{*}\right)$
checking whether $f(r)=x$ and that comply with the compression condition.
The following theorem substantiates these contemplations:

\begin{thm}[No superfluous information]\label{thm:no_superfluous_information}The
shortest feature $f^{\ast}$ of string $x$ and its residual $r$
do not contain much superfluous information
\begin{equation}
K\left(f^{\ast},r\mid x\right)=O\text{\ensuremath{\left(\log l\left(x\right)\right)}}\label{eq:no_sup_info}
\end{equation}
the shortest descriptive map generally does not contain much information

\begin{equation}
l\left(f^{\prime\ast}\right)=O\text{\ensuremath{\left(\log l\left(x\right)\right)}}\label{eq:zero_dm}
\end{equation}
and the description of $x$ is partitioned into $f^{*}$ and $r$:
\begin{equation}
K(x)=K(f^{\ast})+K(r)+O(\log l(x))\label{eq:partition}
\end{equation}
\end{thm}\begin{pf}If $x$ and $l(f^{*})$ is given, then consider
all combinations of strings $g$ with fixed length $l(g)=l(f^{*})$
and strings $q$ with $l(q)<l(x)-l(f^{*})$ of which there are finitely
many. Execute every $(g,q)$-pair as $g(q)$ in parallel, until one
of them prints $x$. We know that some pair will halt and print $x$
at some point, since by assumption we know that there exists a $r$
such that $f^{*}(r)=x$ halts and that pair is part of the executed
set. Let $(\tilde{f},\tilde{r})$ be the first halting pair. Then
the described algorithm implies
\begin{equation}
K\left(\tilde{f},\tilde{r}\mid x,l\left(f^{\ast}\right)\right)=O\text{\ensuremath{\left(1\right)}}
\end{equation}
while the compression condition $l(\tilde{f})+l(\tilde{r})<l(x)$
holds by construction. Therefore, $K\left(\tilde{f},\tilde{r}\mid x\right)\le K\left(l\left(f^{\ast}\right)\right)+O\text{\ensuremath{\left(1\right)}}=O\text{\ensuremath{\left(\log l\left(f^{\ast}\right)\right)}}=O\text{\ensuremath{\left(\log l\left(x\right)\right)}}$.
The boundedness by $O\left(\log l(x)\right)$ is guaranteed, since
$l(f^{*})<l(x)$ by the compression condition.

Now, we exploit Theorem \ref{thm:number_of_features} that entails
that the number of shortest features $d(F)$ is bounded by $2^{O(\log l(x))}$.
In particular, any shortest feature $f^{*}$ can be encoded by an
index $i_{f^{*}}$ bounded by $i_{f^{*}}=O(\log d(F))$. Therefore,
$K\left(f^{*},r\mid x,l\left(f^{\ast}\right),i_{f^{*}}\right)=O\text{\ensuremath{\left(1\right)}}$
which entails the first result (eq.\ (\ref{eq:no_sup_info})):
\begin{equation}
K(f^{*},r\mid x)\le K\left(f^{*},r\mid x,l\left(f^{\ast}\right),i_{f^{*}}\right)+K\left(l\left(f^{\ast}\right)\right)+K\left(i_{f^{*}}\right)+O(1)=O\left(\log l(x)\right)
\end{equation}
by a similar line of reasoning. Let $f^{\prime\ast}$ be shortest
descriptive map corresponding to $f^{*}$. Using Lemma \ref{lem:lemma_feature_length_complexity}
we observe 
\begin{equation}
l\left(f^{\prime\ast}\right)=K\left(r\mid x\right)\leq K\left(f^{*},r\mid x\right)+O(1)=O\text{\ensuremath{\left(\log l\left(x\right)\right)}}
\end{equation}
yielding the second result (eq.\ (\ref{eq:zero_dm})). This result
can be interpreted intuitively since we can take $f^{\prime}$ as
the algorithm described above, with index $i_{f^{*}}$ decoded inside
it, except that it just outputs $r$ instead of $(f,r)$.

Finally, since $f^{*}(r)=x$, on the one hand, $K\left(x\right)\leq K(f^{*},r)+O\left(1\right)\leq K(f^{\ast})+K(r)+O\left(1\right)$.
On the other hand, using Corollary \ref{cor:independence}, we obtain
\[
K(f^{\ast})+K(r)-O(\log l(f^{*}))-O(\log l(r))=K\left(f^{\ast},r\right)+O(1)\leq
\]
\begin{equation}
K\left(f^{\ast},r\mid x\right)+K(x)+O(1)=K\left(x\right)+O\left(\log l(x)\right)
\end{equation}
Since $O(\log l(f^{*}))+O(\log l(r))=O(\log l(x))$ due to the compression
condition, the third result (eq.\ (\ref{eq:partition})) follows.
\qed\end{pf}

We can summarize the results by stating:
\begin{enumerate}
\item $f^{\ast}$ and $r$ do not contain superfluous information: $K\left(f^{\ast},r\mid x\right)\approx0$
\item The shortest descriptive map is very short: $l\left(f^{\prime\ast}\right)\approx0$
\item The description of $x$ breaks down into the separate description
of $f^{\ast}$ and $r$: $K(x)\approx K(f^{\ast})+K(r)$
\end{enumerate}
Finally, Theorem \ref{thm:feature_incompressibility} leads to the
main result of a single compression step: $K(x)\approx l(f^{\ast})+K(r)$.
A compression step consists of taking the shortest feature $f^{\ast}$
and descriptive map $f^{\prime\ast}$ of $x$ according to Definition
\ref{definition1} and computing the residual $r$. The string $x$
is described by the pair $\left(f^{\ast},r\right)$ which has to be
compressed further. However, in this subsection we have derived that
$f^{\ast}$ is both incompressible and independent of $r$. Thus,
we do not have to bother about the feature, we can greedily continue
to compress the residual only. It this is done well, $x$ will also
be compressed well. The following sections explore this compression
scheme.

\subsection{Incremental compression scheme\label{subsec:Incremental-compression-scheme}}

The iterative application of the just described compression step is
called \textit{incremental compression}. Denote $r_{0}\equiv x$ and
let $f_{i}^{\ast}$ be a shortest feature of $r_{i-1}$, $f_{i}^{\prime\ast}$
a shortest (corresponding) descriptive map with $r_{i}=f_{i}^{\prime\ast}\left(r_{i-1}\right)$.
The iteration $i=1,2,\ldots$ continues until some $r_{s}$ is not
compressible (for example, $r_{s}=\epsilon$) any more. This leaves
us with the composition of functions $x=f_{1}^{\ast}\left(f_{2}^{\ast}\left(\cdots f_{s}^{\ast}\left(r_{s}\right)\right)\right)$.

A consequence of such a compression procedure is the pairwise orthogonality
of features obtained this way. After all, if $f_{i}^{*}$ doesn't
know much about $r_{i}$, and $f_{i+1}^{*}$ is part of $r_{i}$,
then $f_{i}^{*}$ doesn't know much about $f_{i+1}^{*}$ either. Formally,
we define the information in $x$ about $y$ as $I(x:y):=K(y)-K(y\mid x)$,
and obtain:

\begin{lem}\label{lem:lemmaxyz}Let $x$, $y$, $z$ be arbitrary
finite strings. Then $I(x:z)\leqslant I(x:y)+K(z\mid y)+O\left(\log l\left(z\right)\right)$.\end{lem}\begin{pf}Using
general properties of the prefix complexity, we expand up to additive
constants
\begin{equation}
K(z)+K(y\mid z,K(z))=K(z,y)=K(z\mid y,K(y))+K(y)\leqslant K(y)+K(z\mid y)
\end{equation}
and also look at the conditional version \citep[eq. (3.22)]{li2009introduction}:
\begin{equation}
K(z\mid x)+K(y\mid z,K(z\mid x),x)=K(z,y\mid x)=K(z\mid y,K(y\mid x),x)+K(y\mid x)\geqslant K(y\mid x)
\end{equation}
We subtract the inequalities and obtain:
\begin{equation}
I(x:z)\leqslant I(x:y)+K(y\mid z,K(z\mid x),x)-K(y\mid z,K(z))+K(z\mid y)
\end{equation}
Observe that $K(y\mid z,K(z\mid x),x)\leqslant K(y\mid z)\leqslant K(y\mid z,K(z))+K(K(z)\mid z)$
(up to additive constants) leading to:

\begin{equation}
I(x:z)\leqslant I(x:y)+K(z\mid y)+K(K(z)\mid z)\leqslant I(x:y)+K(z\mid y)+O\left(\log l\left(z\right)\right).
\end{equation}
\qed\end{pf}

This result can be applied to all features during incremental compression:

\begin{thm}[Orthogonality of features]\label{thm:orthogonality}Let
$x$ be a finite string that is incrementally compressed by a sequence
of shortest features $f_{1}^{*},\ldots,f_{s}^{*}$ that is $x=f_{1}^{\ast}\left(f_{2}^{\ast}\left(\cdots f_{s}^{\ast}\left(r_{s}\right)\right)\right)$.
Then, the features are pairwise \textbf{orthogonal} in terms of the
algorithmic information:
\begin{equation}
I(f_{i}^{*}:f_{j}^{*})=O\left(\lvert i-j\rvert\log l(x)\right)
\end{equation}
 for all $i\neq j$.\end{thm}\begin{pf}First, we prove orthogonality
for the case $j>i$. We denote $r_{i}=(f_{i+1}^{*}\circ f_{i+2}^{*}\circ\ldots\circ f_{j}^{*})(r_{j})$,
where $x=(f_{1}^{*}\circ f_{2}^{*}\circ$$\ldots\circ f_{i}^{*})(r_{i})$.
Using the core idea of Theorem \ref{thm:no_superfluous_information}
we can prove that not much information is required to encode the shortest
features $f_{i+1}^{*},f_{i+2}^{*},\ldots,f_{j}^{*}$ given $r_{i}$.
$f_{j}^{*}$ can be found by iteratively exploiting the algorithm
described in above-mentioned theorem, substituting $r_{m}$ for $x$
and $f_{m+1}^{*}$ for the $n$-th shortest feature $f^{\ast}$, where
$m$ takes values from $i$ to $j-1$. Formally:
\[
K(f_{j}^{*}\mid r_{i})\leqslant K\left(f_{j}^{*}\mid r_{i},l(f_{i+1}^{*}),\ldots,l(f_{j}^{*}),n_{i+1},\ldots,n_{j}\right)+
\]
\begin{equation}
+\overset{j}{\underset{m=i+1}{\sum}}K\left(l(f_{m}^{*})\right)+O\left((j-i)\log l(x)\right)\leqslant O\left((j-i)\log l(x)\right)
\end{equation}
From Corollary \ref{cor:independence}, we know that $I(f_{i}^{*}:r_{i})=O(\log l(f_{i}^{*}))+O(\log l(r_{i}))=O(\log l(x))$
due to the compression condition. We insert this into Lemma \ref{lem:lemmaxyz}
(replacing $x\rightarrow f_{i}^{*}$, $y\rightarrow r_{i}$ and $z\rightarrow f_{j}^{*}$)
to get an upper bound on $I(f_{i}^{*}:f_{j}^{*})$:
\begin{equation}
I(f_{i}^{*}:f_{j}^{*})\leqq I(f_{i}^{*}:r_{i})+K(f_{j}^{*}\mid r_{i})+O\left(\log l(f_{j}^{*})\right)\leq O\left((j-i)\log l(x)\right)
\end{equation}
For the case $j<i$ we first rename $i\leftrightarrow j$: $I(f_{j}^{*}:f_{i}^{*})=O((i-j)\log l(x))$.
We exploit the symmetry of information \citep[Lemma 3.9.2]{li2009introduction}
and obtain:
\begin{align}
I(f_{i}^{*}:f_{j}^{*})\leqslant\log K(f_{i}^{*})+\log K(f_{j}^{*})+2\log\log K(f_{i}^{*})+\\
2\log\log K(f_{j}^{*})+O\left((i-j)\log l(x)\right)=\; & O\left((i-j)\log l(x)\right)
\end{align}
\qed\end{pf}

Finally, we can turn to the interesting question on the optimality
of our compression scheme. One of our main theoretical results shows
that incremental compression finds a description whose length coincides
with the Kolmogorov complexity up to logarithmic terms, i.e.\ achieves
near optimal compression:

\begin{thm}[Optimality of incremental compression]\label{thm:compression_scheme}Let
$x$ be a finite string. Define $r_{0}\equiv x$ and let $f_{i}^{\ast}$
be a shortest feature of $r_{i-1}$ and $r_{i}$ its corresponding
residual description. The compression scheme described above leads
to the description $x=f_{1}^{\ast}\left(f_{2}^{\ast}\left(\cdots f_{s}^{\ast}\left(r_{s}\right)\right)\right)$,
encoded as $D_{s}:=\left\langle s,r_{s},f_{s}^{\ast}\cdots f_{1}^{\ast}\right\rangle $,
for which the following expression holds:
\begin{equation}
K\left(x\right)=\mathbin{\sum_{i=1}^{s}l\left(f_{i}^{\ast}\right)+K\left(r_{s}\right)+O\left(s\cdot\log l(x)\right)}\label{eq:approxK}
\end{equation}
\end{thm}\begin{pf}Iterating the relationship $K(x)=l(f^{*})+K(r)+O\left(\log l(x)\right)$
from Theorem \ref{thm:no_superfluous_information} gives us the required
result, since $l(r_{i})<l(x)$ for all $i=1,\ldots,s$ due to the
iterated compression condition.

We show that $D_{s}:=\left\langle s,r_{s},f_{s}^{\ast}\cdots f_{1}^{\ast}\right\rangle $
is a description of $x$. A program of constant length can take the
number of features $s$, the last residual $r_{s}$ and execute $U\left(\bar{r}_{s}f_{s}^{\ast}\cdots f_{1}^{\ast}\right)$.
By construction, $U\left(\bar{r}_{s}f_{s}^{\ast}\right)\equiv U\left(\left\langle r_{s},f_{s}^{*}\right\rangle \right)$
will halt printing $r_{s-1}$ with the input head at the start of
the remainder $f_{s-1}^{\ast}\cdots f_{1}^{\ast}$. Thus, no self-delimiting
code of the features is necessary. The remainder can be concatenated
to $r_{s-1}$ and executed on $U$ again: $U\left(\bar{r}_{s-1}f_{s-1}^{\ast}\cdots f_{1}^{\ast}\right)$.
This procedure is iterated $s$ times until $x$ is printed.\qed\end{pf}

\subsection{On the number of compression steps\label{subsec:On-the-number}}

The number of compression steps in eq.\ (\ref{eq:approxK}) is in
general difficult to estimate. In the worst case, we could be compressing
merely 1 bit at every step. Then, $s=O\left(l\left(x\right)\right)$
and $l(D_{s})=K\left(x\right)+O\left(l\left(x\right)\cdot\log l\left(x\right)\right)$
which is not satisfactory. Here is an example of a bad case, where
the number of features grows almost linearly:

\begin{rem}\label{rem:remark2}Consider a machine $I_{1}\left(\left\langle \left\langle a,b,y\right\rangle ,\epsilon\right\rangle \right)=y_{1}y_{2}\ldots y_{a-1}1^{b}y_{a}y_{a+1}\ldots y_{n}$,
where $n=l\left(y\right)$. It inserts $b$ ones into string $y$
before the symbol at index $a$. Let $T_{j}$ be the number of this
machine, i.e.\ $U\left(\left\langle \cdot,\left\langle j,\epsilon\right\rangle \right\rangle \right)=T_{j}\left(\left\langle \cdot,\epsilon\right\rangle \right)=I_{1}\left(\left\langle \cdot,\epsilon\right\rangle \right)$.
Define the universal Turing machine $U$ such, that $I_{1}$ is the
first one in the enumeration of all machines: $j=0$. Then $\left\langle j,\epsilon\right\rangle $
will in general be the shortest possible feature. In particular, the
number $j=0$ will be encoded by the empty string, $l\left(\left\langle j,\epsilon\right\rangle \right)=l\left(\bar{\epsilon}\epsilon\right)=l\left(1^{0}0\right)=l\left(0\right)=1$.
The compression condition amounts to
\begin{equation}
\begin{aligned}l\left(\left\langle j,\epsilon\right\rangle \right)+l\left(\left\langle a,b,y\right\rangle \right)= & 1+2l\left(a\right)+1+2l\left(b\right)+1+l\left(y\right)\leq\\
 & l\left(y\right)+2\log_{2}l\left(y\right)+2\log_{2}b+3<\\
 & l\left(y_{1}y_{2}\ldots y_{a-1}1^{b}y_{a}y_{a+1}\ldots y_{n}\right)=l\left(y\right)+b
\end{aligned}
\end{equation}
This condition is fulfilled, if $b\ge c\cdot\log_{2}l\left(y\right)$
for some constant $c$. Thus, if a string of length $n$ contains
at least $c\log_{2}n$ ones, $\left\langle j,\epsilon\right\rangle $
will be the shortest feature.

Consider now a string $z$ of the type $z=1^{a_{1}}01^{a_{2}}0\ldots1^{a_{s}}0$
with $a_{i}\ge c\log_{2}l\left(z\right)$ for all $i$. Then $z$
will have at least $s$ consecutive shortest features $\left\langle j,\epsilon\right\rangle $,
since $z=\left\langle j,\epsilon\right\rangle \left(\left\langle a,b,y\right\rangle \right)$,
where $y$ looks like $z$ except for having one block of ones less.
Thus, we can apply similar arguments to the residual $r_{1}=\left\langle a,b,y\right\rangle $
and obtain $r_{1}=\left\langle j,\epsilon\right\rangle \left(r_{2}\right)$
and so on. Ultimately, the number of features $\left\langle j,\epsilon\right\rangle $
can be bounded from below by an expression proportional to $l\left(z\right)/\log l(z)$.
Interestingly, the overhead produced by various block coding methods
could be reduced in \citep{barmpalias2019compression} by a sophisticated
technique called layered Kraft-Chaitin coding. Its adaptation to incremental
compression is however not straightforward.\end{rem}

In order to bound the overhead $O\left(s\cdot\log l\left(x\right)\right)$
we impose an additional condition on the number of iterations $s$:
we demand that apart from the compression condition, the residual
shall be at least $b$ times smaller than $l(x)$, where $b\ge1$:

\begin{dfn}\label{definition1-b}Let $f$ and $x$ be finite strings.
Denote $D_{f,b}\left(x\right)$ as the set of \textbf{$b$-descriptive
maps} of $x$ given $f$,\textbf{ }the \textbf{$b$-feature }$f$,
\textbf{$b$-residuals} $r$ and\textbf{ $b$-descriptive map} $f'$
similarly to Definition \ref{definition1} adding the condition $l\left(f^{\prime}\left(x\right)\right)\leq\frac{l\left(x\right)}{b}$
to the compression condition: 
\begin{equation}
D_{f,b}\left(x\right)=\left\{ f^{\prime}:\,f\left(f^{\prime}\left(x\right)\right)=x,\,l\left(f\right)+l\left(f^{\prime}\left(x\right)\right)<l\left(x\right),\;l\left(f^{\prime}\left(x\right)\right)\leq\frac{l\left(x\right)}{b}\right\} 
\end{equation}
\end{dfn}

As can be verified by going through the proofs, the most results about
shortest features remain valid for the shortest $b-$features:

\begin{lem}\label{lem:lemmas_and_thms_b_feature}Lemma \ref{lem:lemma_total_feature},
Lemma \ref{lem:lemma_feature_length_complexity}, Theorem \ref{thm:feature_incompressibility},
Lemma \ref{lem:K_f_x}, Lemma \ref{lem:I_f_x}, Theorem \ref{thm:number_of_features},
Theorem \ref{thm:independence-general}, Corollary \ref{cor:independence},
Theorem \ref{thm:no_superfluous_information}, Theorem \ref{thm:orthogonality},
Theorem \ref{thm:compression_scheme} hold for the shortest \textbf{$b$-}feature
and shortest \textbf{$b$-}descriptive map of a finite string $x$,
respectively.\end{lem}\begin{pf}In the following, we will not repeat
all the proofs for $b$-features, but only highlight the parts in
which special care has to be taken.

\paragraph{Lemma \ref{lem:lemma_total_feature}}

The singleton feature $f$ is always a $b$-feature, since its residual
$r=\epsilon$, and thus $0=l\left(r\right)\leq\frac{l(x)}{b}$ holds
trivially for any $b$.

\paragraph{Lemma \ref{lem:lemma_feature_length_complexity}}

To prove the equality $l(f^{*})=K(x\mid r)$ for the shortest $b$-feature
we need to check whether $D_{g,b}\left(x\right)\neq\emptyset$ holds.
But $g^{\prime}=f^{\prime\ast}\in D_{g,b}\left(x\right)$ since $g^{\prime}\in D_{g}\left(x\right)$
by the proof of Lemma \ref{lem:lemma_feature_length_complexity} and
$l\left(g^{\prime}\left(x\right)\right)=l\left(f^{\prime\ast}\left(x\right)\right)=l\left(r\right)\leq l\left(x\right)/b$
since $f^{*}$ is a $b$-feature by assumption of modified Lemma \ref{lem:lemma_feature_length_complexity}.
Using similar arguments in the proof of $l\left(f^{\prime\ast}\right)=K(r\mid x)$,
we show $g^{\prime}\in D_{f^{\ast},b}\left(x\right)$.

\paragraph{Theorem \ref{thm:feature_incompressibility}}

Theorem \ref{thm:feature_incompressibility} is direct consequence
of Lemma \ref{lem:lemma_feature_length_complexity} and general properties
of Kolmogorov complexity.

\paragraph{Lemmas \ref{lem:K_f_x} and \ref{lem:I_f_x}}

Lemmas \ref{lem:K_f_x} and \ref{lem:I_f_x} hold for any features,
and hence hold for $b$-features as well.

\paragraph{Theorem \ref{thm:number_of_features}}

Let us consider the analogue of Theorem \ref{thm:number_of_features}
with shortest $b$-features instead of shortest features. If $l(x)-K(x)\le C$
for some constant $C$ we know from Theorem \ref{thm:number_of_features}
that the number of all features is not greater than $O\left(l(x)^{\alpha}\right)$
for some constant $\alpha$ and $K(f\mid x)\leq O\left(\log l(x)\right)$
for any feature, so this holds for $b$-features as well. If $l(x)-K(x)>C=l\left(f_{0}\right)$
and $K(x)\le l(x)/b$ then the universal feature $f_{0}$ is also
a $b$-feature (since for its residual $l(r)=K(x)\le l(x)/b$). Thus,
similarly to Theorem \ref{thm:number_of_features}, the number of
shortest features is limited by a constant $2^{C+1}-1$ and for them
$K(f^{*}\mid x)=O\left(1\right)$.

Consider the remaining case $K(x)>l(x)/b$ and denote $d=K(x)-\left\lfloor l(x)/b\right\rfloor $,
where $\left\lfloor z\right\rfloor $ is the maximal integer not greater
than the real number $z$. Divide shortest description $g$ of $x$
in two parts $g=g'g''$ where $l\left(g'\right)=\left\lfloor l(x)/b\right\rfloor $
and $l\left(g''\right)=d$. Define some function $h\left(\left\langle y,q\right\rangle \right):=U\left(\left\langle \epsilon,yq\right\rangle \right)$.
Let $j$ be the number of $h$ in the standard enumeration of Turing
machines, $h=T_{j}$. Thus, 
\begin{equation}
U\left(\left\langle g',\left\langle j,g''\right\rangle \right\rangle \right)=T_{j}\left(\left\langle g',g''\right\rangle \right)=h\left(\left\langle g',g''\right\rangle \right)=U\left(\left\langle \epsilon,g'g''\right\rangle \right)=U\left(\left\langle \epsilon,g\right\rangle \right)=x,
\end{equation}
since $g$ is a description of $x$. Therefore, $\hat{f}:=\left\langle j,g''\right\rangle $
is a $b$-feature with its residual $\hat{r}=g'$ if the compression
condition holds: $l\left(\hat{f}\right)+l\left(\hat{r}\right)<l\left(x\right)$,
since the other condition $l\left(\hat{r}\right)\leq l\left(x\right)/b$
holds by construction of $g'$. Let us calculate $l\left(\hat{f}\right)=2l\left(j\right)+1+l\left(g''\right)$
and define a constant $\hat{C}=2l\left(j\right)+1$, so that $l\left(\hat{f}\right)=\hat{C}+d$.
Therefore, $l\left(\hat{f}\right)+l\left(\hat{r}\right)=\hat{C}+d+l\left(g'\right)=\hat{C}+K\left(x\right)-\left\lfloor l(x)/b\right\rfloor +\left\lfloor l(x)/b\right\rfloor =K\left(x\right)+\hat{C}$.
The case $l(x)-K(x)\le\hat{C}$ is similar to the already reviewed
case $l(x)-K(x)\le C$. Thus, consider the case $l(x)-K(x)>\hat{C}$.
Then $l\left(\hat{f}\right)+l\left(\hat{r}\right)=K\left(x\right)+\hat{C}<l(x)-\hat{C}+\hat{C}=l(x)$
and $\hat{f}$ is indeed a $b$-feature of length $l\left(\hat{f}\right)=\hat{C}+d$.

For any shortest $b$-feature $f^{*}$ we have $K(x\mid f^{*})\leq l\left(r\right)+O\left(\log l\left(r\right)\right)\leq\left\lfloor l\left(x\right)/b\right\rfloor +O\left(\log l\left(x\right)\right)$
by Lemma \ref{lem:K_f_x}. Then 
\begin{multline}
I(f^{*}:x)\coloneqq K(x)-K(x\mid f^{*})\geq K\left(x\right)-\left\lfloor l\left(x\right)/b\right\rfloor -O\left(\log l\left(x\right)\right)=\\
d-O\left(\log l\left(x\right)\right)=l\left(\hat{f}\right)-\hat{C}-O\left(\log l\left(x\right)\right)\geq l(f^{*})-O\left(\log l\left(x\right)\right),
\end{multline}
since $l(f^{*})\leq l\left(\hat{f}\right)$ and the constant $\hat{C}$
can be subsumed into $O\left(\log l(x)\right)$. By Lemma \ref{lem:I_f_x}
we conclude $K(f^{*}\mid x)\leq O\left(\log l(x)\right)$ for all
shortest $b$-features $f$ of $x$. Similarly to the proof of Theorem
\ref{thm:number_of_features} the number of shortest $b$-features
is not greater than $O\left(l(x)^{\alpha}\right)$ for some constant
$\alpha$.

\paragraph{Theorem \ref{thm:independence-general} and Corollary \ref{cor:independence}}

In the modified Theorem \ref{thm:independence-general} $\hat{f}$
is a $b$-feature since $l\left(r\right)\leq l(x)/b$ because $f$
is a $b$-feature by assumption of modified Theorem \ref{thm:independence-general}.
The remainder of the proof is analogous. Therefore, Corollary \ref{cor:independence}
is also true for shortest $b$-features.

\paragraph{Theorem \ref{thm:no_superfluous_information}}

We need to consider all combinations of strings $g$ with fixed length
$l(g)=l(f^{*})$ and strings $q$ with $l(q)<l(x)-l(f^{*})$ and $l(q)<l(x)/b$
of which there are finitely many. The remainder of the proof is similar
to the original.

\paragraph{Theorem \ref{thm:orthogonality} and Theorem \ref{thm:compression_scheme}}

Since Theorem \ref{thm:orthogonality} follows from Theorem \ref{thm:no_superfluous_information},
Corollary \ref{cor:independence} and Lemma \ref{lem:lemmaxyz} it
is also true for shortest $b$-features. The same arguments are valid
for Theorem \ref{thm:compression_scheme} which follows from Theorem
\ref{thm:no_superfluous_information}.\qed\end{pf}

Let us fix some factor $b>1$ and search for shortest $b$-features
instead of the shortest features in the compression scheme of Theorem
\ref{thm:compression_scheme}. Applying the compressibility by factor
$b$ to each residual, $l(r_{i})\le l(r_{i-1})/b$, leads to $l(r_{s-1})\le l(x)/b^{s-1}$.
We might continue this process until $r_{s}$ is not compressible.
Since $r_{s-1}$ is compressible it cannot be the empty string $\epsilon$,
hence $l(r_{s-1})\geq1$ and we obtain the bound $s\le\log_{b}\left(l(x)/l(r_{s-1})\right)+1\le\log_{b}l(x)+1=O\left(\log l\left(x\right)\right)$.
In this way, the estimation error $O\left(s\cdot\log l\left(x\right)\right)$
in Theorem \ref{thm:compression_scheme} becomes $O\left((\log l\left(x\right))^{2}\right)$,
which is quite small. Note, that there is no problem with the asymptotics
since $b$ is strictly larger than 1 and is fixed. Further, this processes
is always possible since the singleton feature always exists for compressible
strings and it is a $b$-feature at the same time.

\subsection{A special case: well-compressible strings\label{subsec:well-compressible}}

If we are aiming for a practical compression algorithm, it is reasonable
to assume that the strings actually are compressible. In this section,
we would like to take a look at this special case. As we shall see,
the results of the theory can be substantially strengthened this way.
In particular, the bounds on the lengths of the shortest feature and
descriptive maps will turn out to be constant.

Let $x$ be compressible by a factor $b$: $K(x)\le l(x)/b$, $b>1$.
Clearly, by Lemma \ref{lem:lemma_universal_feature} a universal feature
exists if $x$ is long enough. Then the shortest feature is bounded
by the length of the universal feature: a constant. Conversely, if
$x$ is not long enough, say $l(x)<a$, then the length of the shortest
features is also bounded by a constant, since $l(f)<l(x)<a$ by the
compression condition. This is substantiated by

\begin{thm}\label{thm:thm_short_features}Let $x$ be a $b$-compressible
string for some fixed factor $b>1$. Then the length of any shortest
feature $f^{\ast}$ of $x$ is bounded, $l\left(f^{\ast}\right)=O\left(1\right)$.
More precisely, $l\left(f^{\ast}\right)\leq\max\left\{ C,\frac{C}{b-1}\right\} =:C_{0}$,
where $C:=l\left(f_{0}\right)$ is the length of the universal feature
$f_{0}$. The same statement holds for the shortest $b$-feature.
Moreover, if additionally $l\left(x\right)>\frac{Cb}{b-1}$ then $f_{0}$
is both a feature and a $b$-feature of $x$, and $l\left(f^{\ast}\right)\leq C$.\end{thm}\begin{pf}After
Lemma \ref{lem:lemma_universal_feature}, the residual $r$ of the
universal feature $f_{0}$ is the shortest description: $l(r)=K(x)$.
$f_{0}$ becomes a feature of $x$ if the compression condition $l(f_{0})+l(r)=C+K(x)<l(x)$
is fulfilled, which is the case if $x$ is long enough: $l\left(x\right)>\frac{Cb}{b-1}$.
After all, since $x$ is $b$-compressible,
\begin{equation}
l(f_{0})+l(r)=C+K(x)\le C+\frac{l\left(x\right)}{b}<l(x)\frac{b-1}{b}+l(x)\frac{1}{b}=l(x)
\end{equation}
Note that $f_{0}$ is also a $b$-feature, since $l(r)=K(x)\le l(x)/b$.
Let $f^{*}$ be a shortest feature $x$. Then, $l\left(f^{\ast}\right)\leq l\left(f_{0}\right)=C$.
Since $f_{0}$ is a $b$-feature, the length of a shortest $b$-feature
is also bounded by $C$. Conversely, if $l(x)$ is not long enough,
using Lemma \ref{lem:lemma_total_feature} $l\left(f^{*}\right)\le K\left(x\right)\le l\left(x\right)/b\leq\frac{C}{b-1}$.
In both cases, $l(f^{*})$ is bounded by a constant for both shortest
feature and shortest $b$-feature.\qed\end{pf}

Note that if $b\geq2$ then $\frac{C}{b-1}\le C$, so the length of
shortest feature of a $b$-compressible string is always limited by
$C=l\left(f_{0}\right)$. This theorem implies that if the $b$-compressibility
assumption holds, we do not require $x$ to be sufficiently long.
The existence of a \textit{short} feature, i.e.\ $l(f)\le C_{0}$,
is guaranteed. Therefore, many of the above theorems simplify considerably.
We obtain not just $K(f^{\ast}\mid x)\approx0$ (Theorem \ref{thm:no_superfluous_information}),
but a much stronger proposition $K(f^{\ast})=O\left(1\right)$ and
$l(f^{\ast})=O\left(1\right)$ in case of $b$-compressible $x$.
This circumstance demonstrates that features can be very short. Incompressibility
of $f^{*}$ (Theorem \ref{thm:feature_incompressibility}) and independence
between $f^{*}$ and $r$ (Corollary \ref{cor:independence}) follow
trivially. The number of shortest features (Theorem \ref{thm:number_of_features})
is also bounded by a constant $2^{C_{0}}$. It turns out that the
shortest descriptive map is also short:

\begin{thm}[Short descriptive maps, no superfluous information in short features]\label{thm:no_superfluous_information_short}Let
$f$ be a short feature of a compressible $x$, hence $f\left(r\right)=x$,
$l\left(f\right)+l\left(r\right)<l\left(x\right)$, $l(f)=O\left(1\right)$.
Then there exists a residual $q$ such that $f\left(q\right)=x$,
$l\left(f\right)+l\left(q\right)<l\left(x\right)$ and $K\left(q\mid x\right)=O\left(1\right)$.
If $f^{\prime\ast}$ is a shortest descriptive map given $f$, then
$K(f^{\prime\ast}(x)\mid x)=l\left(f^{\prime\ast}\right)=O\left(1\right)$.\end{thm}\begin{pf}Since
$f$ is a feature, there is a residual $r$ fulfilling $f(r)=x$ and
$l(r)<l(x)-l(f)$. Consider all strings shorter than $l(x)-l(f)$
and execute them in parallel on $f$. Denote the algorithm performing
this with $S(f,x)$. This algorithm will halt at some point, since
$r$ is among the executed strings. Let $q$ be the first string that
prints $x$. Then $S(f,x)=q$ and $K(q\mid f,x)=O(1)$ since $S$
is an algorithm of constant length. Encode both $S$ and $f$ into
a descriptive map $g'$ operating like $g'(x):=S(f,x)=q$. Since $l(f)=O(1)$,
we conclude $l(g')=O(1)$ as well and $K\left(q\mid x\right)\leq l\left(g'\right)=O\left(1\right)$.
Now define $f^{\prime\ast}$ as a shortest descriptive map, that is
$f^{\prime\ast}\in D_{f}$ and $l\left(f^{\prime\ast}\right)$ is
minimal. Then $l\left(f^{\prime\ast}\right)\leq l\left(g'\right)=O\left(1\right)$
and $K(f^{\prime\ast}(x)\mid x)=l\left(f^{\prime\ast}\right)=O\left(1\right)$.\qed\end{pf}

\subsection{Incremental compression scheme for $b$-compressible strings}

What are the implications for the whole compression scheme? Suppose
$x$ is a $b$-compressible string and $l\left(x\right)>\frac{Cb}{b-1}$.
Denote $r_{0}\equiv x$ and start an iterative process of compression:
let $f_{i+1}^{\ast}$ be a shortest $b$-feature of $r_{i}$, $f_{i+1}^{\prime\ast}$
a shortest corresponding descriptive map and $r_{i+1}=f_{i+1}^{\prime\ast}\left(r_{i}\right)$.
We continue this process until either $l\left(r_{i}\right)\leq\frac{Cb}{b-1}$
or $r_{i}$ is not $b$-compressible for some $i$. Denote this $i$
by $s$. Just like in Sect.\ \ref{subsec:Incremental-compression-scheme}
we obtain $x=f_{1}^{\ast}\left(f_{2}^{\ast}\left(\cdots f_{s}^{\ast}\left(r_{s}\right)\right)\right)$
and $D_{s}:=\left\langle s,r_{s},f_{s}^{\ast}\cdots f_{1}^{\ast}\right\rangle $
is a description of $x$.

\begin{thm}\label{thm:b_comp_sheme}Given the compression scheme
above the following relationships hold:
\begin{eqnarray}
K\left(x\right)-O\left(1\right)\leq l\left(D_{s}\right) & \leq & K\left(x\right)b+O\left(\log l\left(r_{s}\right)\right)+O\left(s\right)\leq K\left(x\right)b+O\left(\log l(x)\right)\\
l\left(f_{i}^{\ast}\right) & \leq & C=l(f_{0})\quad\mbox{for }i=1,\ldots,s\\
s & = & O\left(\log l(x)\right)
\end{eqnarray}
\end{thm}\begin{pf}Since $D_{s}:=\left\langle s,r_{s},f_{s}^{\ast}\cdots f_{1}^{\ast}\right\rangle $
is a description of $x$, we get $l\left(D_{s}\right)\geq K\left(x\right)-O\left(1\right)$.
Since $l\left(f_{i}^{\ast}\right)\leq C$ by Theorem \ref{thm:thm_short_features},
we have
\begin{equation}
l(D_{s})=l\left(\left\langle s,r_{s},f_{s}^{\ast}\cdots f_{1}^{\ast}\right\rangle \right)=l\left(r_{s}\right)+O\left(\log l\left(r_{s}\right)\right)+O\left(s\right)
\end{equation}
In the first case, $l\left(r_{s}\right)\leq\frac{Cb}{b-1}$. Then
$l\left(D_{s}\right)=O\left(s\right)$, since $\frac{Cb}{b-1}$ is
a constant independent of $x$. In the second case, $r_{s}$ is not
$b$-compressible, ergo $l\left(r_{s}\right)\leq K\left(r_{s}\right)b\leq\left(K\left(x\right)+O\left(s\right)\right)b=K\left(x\right)b+O\left(s\right)$,
since $r_{s}$ can be computed from $x$ by a sequence of descriptive
maps, each of which is bounded by a constant (so $K\left(r_{i+1}\right)\leq K\left(r_{i}\right)+O\left(1\right)$
and $K\left(r_{s}\right)=K\left(x\right)+O\left(s\right)$). It follows:
$l\left(D_{s}\right)\leq K\left(x\right)b+O\left(\log l\left(r_{s}\right)\right)+O\left(s\right)\leq K\left(x\right)b+O\left(\log l\left(x\right)\right)+O\left(s\right)$.

The bound on $s$ is derived from the application of the factor $b$
to each residual, $l(r_{i})\le l(r_{i-1})/b$, leading to $l(r_{s-1})\le l(x)/b^{s-1}$.
Since $r_{s-1}$ is compressible it cannot be the empty string $\epsilon$,
hence $l(r_{s-1})\geq1$ and we obtain the bound $s\le\log_{b}\left(l(x)/l(r_{s-1})\right)+1\le\log_{b}l(x)+1=O\left(\log l(x)\right)$.\qed\end{pf}

Note that the constant in the $O$-notation here depends on $b$ but
not on $x$. If $b$ is close to $1$ then $l\left(D_{s}\right)$
will be close to $K\left(x\right)$, making $D_{s}$ a quite short
description of $x$.

\subsection{Comparison of compression schemes\label{subsec:Comparison-of-compression}}

\begin{table}[h]
\begin{tabular*}{1\textwidth}{@{\extracolsep{\fill}}|>{\centering}m{0.22\textwidth}|>{\centering}m{0.22\textwidth}|>{\centering}m{0.22\textwidth}|>{\centering}m{0.22\textwidth}|}
\hline 
 & Plain incremental compression & Using $b$-features & Using $b$-features,

early termination\tabularnewline
\hline 
\hline 
Compression condition & $l\left(f_{i+1}^{\ast}\right)+l\left(r_{i+1}\right)<l\left(r_{i}\right)$ & $l\left(f_{i+1}^{\ast}\right)+l\left(r_{i+1}\right)<l\left(r_{i}\right)$

and $l\left(r_{i+1}\right)\le l\left(r_{i}\right)/b$ & $l\left(f_{i+1}^{\ast}\right)+l\left(r_{i+1}\right)<l\left(r_{i}\right)$

and $l\left(r_{i+1}\right)\le l\left(r_{i}\right)/b$\tabularnewline
\hline 
Halting condition & incompressibility:

$K\left(r_{i}\right)\geq l\left(r_{i}\right)$ & incompressibility:

$K\left(r_{i}\right)\geq l\left(r_{i}\right)$ & $l\left(r_{i}\right)\leq\frac{Cb}{b-1}$ or

$K\left(r_{i}\right)>l\left(r_{i}\right)/b$\tabularnewline
\hline 
Number of steps (worst case) & $O\left(l\left(x\right)\right)$ & $O\left(\log l(x)\right)$ & $O\left(\log l(x)\right)$\tabularnewline
\hline 
Overhead on each step $l\left(f_{i+1}^{\ast}\right)+K\left(r_{i+1}\right)-K\left(r_{i}\right)$ & $O\left(\log l(r_{i})\right)$ $\leq O\left(\log l(x)\right)$ & $O\left(\log l(r_{i})\right)$ $\leq O\left(\log l(x)\right)$ & $O\left(1\right)$\tabularnewline
\hline 
Length of description $D_{s}:=\left\langle s,r_{s},f_{s}^{\ast}\cdots f_{1}^{\ast}\right\rangle $ & $K\left(x\right)+O\left(l(x)\log l(x)\right)$ & $K\left(x\right)+O\left(\left(\log l(x)\right)^{2}\right)$ & $\leq K\left(x\right)b+O\left(\log l(x)\right)$\tabularnewline
\hline 
\end{tabular*}

\caption{\label{tab:Comparison-of-compression}Comparison of compression schemes}
\end{table}

We present three incremental compression schemes, summarized in Table
\ref{tab:Comparison-of-compression}. All three are based on a greedy
selection of the shortest feature or $b$-feature, which leads to
a decomposition of information into an incompressible feature and
a residual information that is to be compressed further. The first
scheme does not possess any free parameters and the absolutely shortest
feature is searched for. The applicability condition of this scheme
is the existence of a feature per se, i.e.\ the compressibility of
initial data $r_{0}\equiv x$ or the current residual $r_{i}$. As
long as the current residual $r_{i}$ fulfills the inequality $l(r_{i})-K(r_{i})>C=l(f_{0})$,
all shortest features $f_{i+1}^{*}$ will be short: $l\left(f_{i+1}^{*}\right)\le C$
and the overhead description at this step will be small: $l\left(f_{i+1}^{\ast}\right)+K\left(r_{i+1}\right)-K\left(r_{i}\right)=O(1)$.
In the end, however, when the condition $l(r_{i})-K(r_{i})>C$ is
violated, the boundedness of feature lengths is not guaranteed any
more, and the overhead costs become logarithmic $l\left(f_{i+1}^{\ast}\right)+K\left(r_{i+1}\right)-K\left(r_{i}\right)=O\left(\log l(r_{i})\right)\leq O\left(\log l(x)\right)$.
The main problem of this scheme is the lack of an estimate of the
number of compression steps $s$. In the worse case, each subsequent
$r_{i}$ is compressed by merely $O(1)$ bit, which leads to $s=O\left(l\left(x\right)\right)$
compression steps making the estimate in eq.\ (\ref{eq:approxK})
unsatisfactory.

In order to limit the number of steps we have introduced a second
scheme in which it is necessary to fix a number $b>1$ that determines
the minimal compression of the residual on each step. The applicability
condition of this scheme is the existence of a $b$-feature of the
current residual $r_{i}$, i.e.\ the just mentioned compression condition
$l(r_{i+1})\le l(r_{i})/b$ on the current residual. As long as the
conditions $l\left(r_{i}\right)>\frac{Cb}{b-1}$ and $K\left(r_{i}\right)\le l\left(r_{i}\right)/b$
are fulfilled all shortest $b-$features $f_{i+1}^{\ast}$ will be
short: $l\left(f_{i+1}^{\ast}\right)\le C_{0}$ (Theorem \ref{thm:thm_short_features})
and the overhead at this step will be small: $l\left(f_{i+1}^{\ast}\right)+K\left(r_{i+1}\right)-K\left(r_{i}\right)=O(1)$.
Ultimately, however, this bound on the length can not be guaranteed
any more, and the overhead bounds become logarithmic. The main difference
to the first scheme is that now the number of steps is bounded by
$\log_{b}\left(l(x)\right)+1$, allowing us to regulate the maximum
number of steps by changing $b$. It is due to this bound that the
discrepancy between our final description and $K(x)$ does not differ
by more than $O\left(\left(\log l(x)\right)^{2}\right)$.

The third scheme is analogous to the second, but we stop at an earlier
step as soon as there is no short $b-$feature any more. Since up
to this point the overhead is constant, the overall estimation error
is merely $O\left(\log l(x)\right)$. However, the last residual $r_{s}$
may remain compressible, albeit not by more than factor $b$.

\section{Computable incremental compression\label{sec:Computable-IC}}

The schemes presented in the previous subsection would be viable compression
algorithms if they were actually computable. Unfortunately, they inherit
the incomputability of the Kolmogorov complexity, which boils down
to the halting problem. It is well-known that the Kolmogorov complexity
is not only incomputable, but it can not even be approximated to any
precision \citep[follows from][Theorem 2.3.2]{li2009introduction}.
The reason why we have found an approximation to $K(x)$ is that our
own algorithm is incomputable, due to the incomputability of the shortest
features and the impossibility to effectively check for the incompressibility
of the residual.

In this section we present a computable version of incremental compression.
The main motivation of our approach is to be able to compress data
faster than non-incremental methods, since even by exhaustive search
a set of short pieces can be expected to be easier to find than the
full description at once. Indeed, if we neglect execution time, an
exhaustive search for a program of length $K(x)$ takes about $T_{\mbox{non-incremental}}:=2^{K(x)}$
time steps. Inserting our result from eq.\ (\ref{eq:approxK}) leads
to
\begin{equation}
T_{\mbox{non-incremental}}\approx2^{K(r_{s})}\prod_{i=1}^{s}2^{l(f_{i}^{*})}
\end{equation}
In contrast to that, searching the shortest features incrementally
can be done in a greedy fashion: at every compression step we look
for the shortest feature and can be certain to approximate the Kolmogorov
complexity of the string well, without the necessity of backtracking.
Since according to eq.\ (\ref{eq:zero_dm}) the length of the shortest
descriptive map is small, incremental search is expected to take merely
\begin{equation}
T_{\mbox{incremental}}\approx2^{K(r_{s})}+\sum_{i=1}^{s}2^{l(f_{i}^{*})}\ll T_{\mbox{non-incremental}}\label{eq:eff_boost}
\end{equation}
time steps. This chain of reasoning is the basis why we can expect
incremental compression to be faster. Nevertheless, we can not really
neglect the execution time and take it into account in the following.

In this section, we introduce two versions of computable incremental
compression algorithms, Greedy-ALICE and ALICE. The latter is the
general version for which we derive the time complexity.

\subsection{The algorithms Greedy-ALICE and ALICE}

We define a new universal Turing machine $W$, which takes inputs
of the following form: $W\left(\left\langle 2,x,a\right\rangle \right)$,
where $x$ is the string to be compressed, $a=f^{\prime}f$ is an
autoencoder consisting of a descriptive map and a feature. Let $W$'s
outputs be of the form $\left\langle y,r,f\right\rangle $ with $r:=f^{\prime}(x)$
and $y:=f(r)$. Such strange output is needed in order to check whether
$f(f^{\prime}(x))=x$, compute $r$ and extract $f$ from $a$ in
algorithms \ref{algo:greedyalice} and \ref{algo:alice} without wasting
any time. The time of execution $W\left(\left\langle 2,x,a\right\rangle \right)$
is equal to $t=t_{1}+t_{1}^{\prime}+O\left(l\left(x\right)\right)+O\left(l\left(r\right)\right)$,
where $t_{1}$ is the time of execution of $f(r)$ and $t_{1}^{\prime}$
is the time of execution of $f^{\prime}(x)$, and additional terms
include time of comparing $f(f^{\prime}(x))$ and $x$, passing result
of $f^{\prime}(x)$ to $f$, etc. Since $f(r)$ and $f^{\prime}(x)$
have to output $x$ and $r$, respectively, and print it somewhere,
$l\left(r\right)=O\left(t_{1}\right),l\left(x\right)=O\left(t_{1}^{\prime}\right)$.
Thus, $t=O\left(t_{1}+t_{1}^{\prime}\right)$.

If we eventually find some description $x=\left(f_{1}\circ f_{2}\ldots\circ f_{s}\right)\left(r_{s}\right)$,
it is strictly a description of $x$ on another machine $V$ whose
input has the form $D_{s}:=\left\langle s,r_{s},f_{s}f_{s-1}\cdots f_{1}\right\rangle $
and which outputs $x$. Since $V$ runs on a prefix code, it can read
$s$ and $r_{s}$ and then run the initial universal Turing machine
$U$ on the remaining string. After it reads exactly $f_{s}$, $U$
halts because $f_{i}(r_{i})=r_{i-1}$ by construction (see also proof
of Theorem \ref{thm:compression_scheme}). In order to search for
autoencoders in parallel, the algorithms \ref{algo:greedyalice} and
\ref{algo:alice} below dovetail the computation. All self-delimiting
autoencoders $a$ of length less than $i$ are run as part of $W$'s
input for $2^{i-l(a)}$ steps in phase $i$, $i=1,2,\ldots,\infty$,
until some program prints $x$ and halts (compare with algorithm SEARCH
\citep[Theorem 7.5.1]{li2009introduction}).

\SetEndCharOfAlgoLine{}
\begin{algorithm}[H] 
\label{algo:greedyalice}
\KwData{$x$ - finite string, $W$ - universal Turing machine, $\mathcal{F}$ - list of features as a global variable, $T_I$ - runtime, $r$ - last residual as a global variable}     
\SetKwProg{Fn}{Function}{}{}

\Fn{GreedyALICE($x$)}{
	run GreedySearch($x$) for $T_I$ steps (in terms of the execution of $W$) \\
	return $\mathcal{F}, r$	(global variables)
}

\Fn{GreedySearch($x$)}{
	$\mathcal{F} \leftarrow \varnothing$\;
	$r \leftarrow x$\;
	\While{True}{
		$r, f$ = SearchAutoencoder($x$)\;
		Append $f$ to $\mathcal{F}$\;
		$x\leftarrow r$\;
	}
}

\Fn{SearchAutoencoder(x)}{
	\For{$i=1,2,\ldots,\infty$}{
		\For{$a$ in $B^*$, $l(a)<i$}{
			\For {$2^{i-l(a)}$ steps}{
				run $W(\langle2,x,a\rangle)$ for 1 step returning $y,r,f$ if halts\;
				\If{$W$ halts and $y=x$ and $l(f)+l(r)<l(x)$}{
					return $r,f$
				}
			}
		}
	}
}

\caption{Greedy ALgorithm for Incremental ComprEssion (Greedy-ALICE)}
\end{algorithm}

ALICE (Alg.\ \ref{algo:alice}) generalizes Greedy-ALICE by search
for all possible compositions of functions. The search can be visualized
as a tree, which root node is $\left(f=\varnothing,f^{\prime}=\varnothing,x\right)$
and all other nodes have form $(f,f^{\prime},r)$ such that $r_{\mbox{parent}}=f\left(r_{\mbox{child}}\right)$,
$r_{\mbox{child}}=f^{\prime}\left(r_{\mbox{parent}}\right)$. With
time, the tree grows in width and in depth as each completed and successful
computation of $f\left(f^{\prime}(r)\right)=r$ starts a self-similar
computation of descriptions of $f^{\prime}(r)$. In the first level
of this search tree all possible autoencoders $a_{1}=f_{1}'f_{1}$
are considered such that the fraction $2^{-l\left(a_{1}\right)}$
of the total computation time is allocated at autoencoder $a_{1}$.
In the beginning this time fraction is spent on the execution of $W\left(\left\langle 2,x,a_{1}\right\rangle \right)$.
If $a_{1}$ indeed reconstructs $x$ and fulfills the compression
condition $l\left(f_{1}\right)+l\left(r_{1}\right)\leq l\left(x\right)$
we obtain a residual $r_{1}$. The remaining time of that fraction
is spent searching for a description of $r_{1}$. This is the way
the search time branches and the time given for this branch is again
distributed among the various autoencoders $a_{2}=f_{2}^{\prime}f_{2}$
on the second level and so on.

\SetEndCharOfAlgoLine{}
\begin{algorithm}[H] 
\label{algo:alice}
\KwData{$x$ - finite string, $W$ - universal Turing machine, $D=\varnothing$ - list of descriptions of $x$ as a global variable, $T_I$ - runtime}   
\SetKwProg{Fn}{Function}{}{}

\Fn{ALICE(x)}{
	run SearchAutoencoderRecursively($\langle x, \varnothing\rangle$) for $T_I$ steps (in terms of the execution of $W$)
}

\Fn{SearchAutoencoderRecursively($\langle x, \mathcal{F}\rangle$)}{
	$status \leftarrow$ dictionary, valued UNHALTED by default, for keeping track of the status of program $a$\\
	\For{$i=1,2,\ldots,\infty$}{
		\For{$a$ in $B^*$, $l(a)<i$}{
			\For {$2^{i-l(a)}$ steps}{
				\If{$status[a] = $ UNHALTED}{
					run $W(\langle2,x,a\rangle)$ for 1 step returning $y,r,f$ if halts\;
					\If{W halts}{
						$status[a]\leftarrow$ HALTED\\
						\If{$y=x$ and $l(f)+l(r)<l(x)$}{
							$\tilde{\mathcal{F}}\leftarrow f$ prepend to $\mathcal{F}$\\
							append $\langle|\tilde{\mathcal{F}}|,r,\tilde{\mathcal{F}}\rangle$ to $D$\\  
							$status[a] \leftarrow \langle r,\tilde{\mathcal{F}}\rangle$\;
							\textbf{continue}\;
						}
					}
				}
				\If{$status[a]\neq$ UNHALTED and $status[a]\neq$ HALTED}{
					run SearchAutoencoderRecursively($status[a]$) for 1 step
				}
			}
		}
	}
}

\caption{ALgorithm for Incremental ComprEssion (ALICE)}
\end{algorithm}

\subsection{Bounds on running time}

We now derive an expression for the total computation time needed
to compute a description of the form $\left\langle s,r_{s},f_{s}f_{s-1}\ldots f_{1}\right\rangle $.
First, we derive the general time allocation in ALICE. In order to
do so, we introduce some abstract process $X$, which runs on node
assigned to autoencoder $a$ (informally speaking, it can be anything
which happens under ``for $2^{i-l(a)}$ steps do'' in Algorithm~\ref{algo:alice}).
$X$ can be, for instance, the computation of $W\left(\left\langle 2,x,a\right\rangle \right)$,
the storage of a generated description in $D$ and then search of
some depth-1 description of a newly generated residual $r=f^{\prime}(x)$.

\begin{lem}\label{lem:parallel_computation} If ALICE computes $\tau$
steps on a node assigned to autoencoder $a$, the total computation
time of ALICE is between $2^{l(a)-l\left(a_{\min}\right)-1}\tau$
and $2^{l(a)+1}\tau$ steps, where $a_{\min}$ is the shortest codeword
in the prefix code. \end{lem}

\begin{pf}According to the algorithm, on each iteration $i\geq l(a)$
precisely $2^{i-l(a)}$ steps are allotted for the computation of
node $a$. Let $t(n):=\stackrel[i=l(a)]{n}{\sum}2^{i-l(a)}=2^{n-l(a)+1}-1$
be total time allocated for node $a$ until the $n$-th iteration.
On some value of $n$ we will obtain $t(n-1)<\tau\le t(n)$ and we
obtain $2^{n-l(a)}-1<\tau\le2^{n-l(a)+1}-1$ and thus $n-l(a)\le\log\tau<n-l(a)+1$
which can be turned around as
\begin{equation}
l(a)+\log\tau-1<n\le l(a)+\log\tau\label{eq:tau_a_n}
\end{equation}
If for some other string $b\neq a$ the computation of $W\left(\left\langle 2,x,b\right\rangle \right)$
halts earlier, ALICE continues with the next incremental step and
node, assigned to $b$, keeps being occupied indefinitely. Thus, the
total computation time of ALICE until this moment is strictly $T=\sum_{i=1}^{n}2^{i}\theta_{i}$
with $\theta_{i}:=\underset{a':l(a')\leq i}{\sum}2^{-l(a')}$. This
complies with the bounds 
\begin{equation}
2^{n}\theta_{n}\leq T\leq2^{n+1}\label{eq:bounds_on_T}
\end{equation}
where the upper bound is proven in \citep[Theorem 7.5.1]{li2009introduction}.
Using the ineq.\ (\ref{eq:tau_a_n}) on $n$, we obtain $n>l(a)+\log\tau-1\ge l(a)-1\geq l\left(a_{\min}\right)-1$.
In particular, $l\left(a_{\min}\right)\le n$ and therefore $a_{\min}$
participates in $\theta_{n}$ no matter what $a$ we consider, leading
to $2^{-l\left(a_{\min}\right)}\le\theta_{n}$. Inserting it into
ineqs.\ (\ref{eq:bounds_on_T}) and (\ref{eq:tau_a_n}) we obtain

\[
2^{l(a)-l(a_{\min})-1}\tau<2^{n-l(a_{\min})}\leq T\leq2^{n+1}\le2^{l(a)+1}\tau.
\]
\qed\end{pf}

Since $l\left(a_{\min}\right)$ does not depend on compressed string
we shall consider it as a constant. Then lower and upper bounds are
equal up to multiplicative constant and the total computing time is
$O\left(\text{\ensuremath{\tau}}2^{l(a)}\right)$. Now we extend this
reasoning from a single autoencoder to the search for all compositions
of functions in ALICE.

\begin{thm}\label{thm: parallel_total_time-1}ALICE finds the autoencoders
$\left(f_{1}^{\prime},f_{1}\right),\ldots,\left(f_{s}^{\prime},f_{s}\right)$
and a description $\left\langle s,r_{s},f_{s}f_{s-1}\ldots f_{1}\right\rangle $
of $x$ in time
\[
T_{I}=O\left(\stackrel[i=1]{s}{\sum}\left(t_{i}+t_{i}^{\prime}\right)2^{O(i)+\stackrel[k=1]{i}{\sum}l(f_{k})+l(f_{k}^{\prime})}\right)
\]
 where $t_{i}$ and $t_{i}^{\prime}$ is time to compute $f_{i}\left(r_{i}\right)$
and $f_{i}^{\prime}\left(r_{i-1}\right)$, respectively.\end{thm}

\begin{pf}Consider the base case $s=1$ of induction. The computing
branch labed by autoencoder $a_{1}$ can be doing the following: ``compute
$W\left(\left\langle 2,x,a\right\rangle \right)$, check $y=x$ and
the compression condition'' for time $t=O\left(t_{1}+t_{1}^{\prime}\right)$.
Using Lemma \ref{lem:parallel_computation} the total search time
for a depth-1 description of $x$ is $O\left(t2^{l\left(a_{1}\right)}\right).$
Therefore, $T_{I}=t2^{l(a_{1})+O(1)}=O\left(\left(t_{1}+t_{1}^{\prime}\right)2^{l(f_{1})+l(f_{1}^{\prime})+O(1)}\right)$
and the base case is fulfilled.

Computing the depth-$(s+1)$ description of the form $x=\left(f_{1}\circ\cdots\circ f_{s+1}\right)\left(r_{s+1}\right)$
can be regarded as computing a depth-1 description of $x$ $\left\langle 1,r_{1},f_{1}\right\rangle $
and then computing a depth-$s$ description of $r_{1}$ on the same
branch, assigned to $a_{1}=f_{1}^{\prime}f_{1}.$ The computation
process could then be described as ``compute $W\left(\left\langle 2,x,a_{1}\right\rangle \right)$
and then find a depth-$s$ description of $r_{1}$''. Since $x=f_{1}\left(r_{1}\right)$
and $r_{1}=\left(f_{2}\circ\cdots\circ f_{s+1}\right)\left(r_{s+1}\right)$,
assuming the induction hypothesis for $s$, the computation time becomes

\[
\tau=O\left(t_{1}+t_{1}^{\prime}\right)+O\left(\stackrel[i=1]{s}{\sum}\left(t_{i+1}+t_{i+1}^{\prime}\right)2^{O(i)+\stackrel[k=1]{i}{\sum}l\left(f_{k+1}\right)+l\left(f_{k+1}^{\prime}\right)}\right)
\]
Deploying Lemma \ref{lem:parallel_computation}, we obtain the total
computation time $T_{I}$ as
\begin{equation}
\begin{aligned}T_{I}= & \tau2^{l(a_{1})+O(1)}=O\left(2^{l(f_{1})+l(f_{1}^{\prime})+O(1)}\left[t_{1}+t_{1}^{\prime}+\stackrel[i=1]{s}{\sum}\left(t_{i+1}+t_{i+1}^{\prime}\right)2^{O(i)+\stackrel[k=1]{i}{\sum}l\left(f_{k+1}\right)+l\left(f_{k+1}^{\prime}\right)}\right]\right)=\\
 & O\left(2^{l(f_{1})+l(f_{1}^{\prime})+O(1)}\left[t_{1}+t_{1}^{\prime}+\stackrel[i=2]{s+1}{\sum}\left(t_{i}+t_{i}^{\prime}\right)2^{O(i-1)+\stackrel[k=2]{i}{\sum}l\left(f_{k}\right)+l\left(f_{k}^{\prime}\right)}\right]\right)=\\
 & O\left(\stackrel[i=1]{s+1}{\sum}\left(t_{i}+t_{i}^{\prime}\right)2^{O(i)+\stackrel[k=1]{i}{\sum}l(f_{k})+l(f_{k}^{\prime})}\right)
\end{aligned}
\end{equation}
We see that error term $O(i)$ is expained by accumulation of $O(1)$
error term from Lemma \ref{lem:parallel_computation}. We get exactly
the same formula of $T_{I}$ for the depth-$(s+1)$ description and
the theorem is proven by induction.\qed\end{pf}

\subsection{Computable incremental compression in practice}

Given such promising theoretical prospects, one might wonder how ALICE
might work in pratice. Indeed, some practical success can be claimed
by WILLIAM -- our Python-based implementation of incremental compression
\citep{franzetal2018,franz2019agi}. In a nutshell, using an algorithm
similar to ALICE, WILLIAM enumerates all autoencoders $(f,f')$ expressed
as Python abstract syntax trees sorted by the sum of their description
lengths $l(f)+l(f')$ until it finds an autoencoder complying with
the compression condition. WILLIAM demonstrates that much deeper trees
can be found efficiently than we could hope to find in a non-incremental
fashion. It is able to solve a diverse set of tasks including the
compression and prediction of simple sequences, recognition of geometric
shapes, write simple code based on test cases, self-improve by solving
some of its own problems and play tic-tac-toe when attached to a the
universally intelligent agent AIXI \citep{Hutter04uaibook} without
being specifically programmed for that game. Recent results include
the emergence of simple versions of various machine learning methods,
such as linear regression and classification, data centralization
and decision trees, as special cases of incremental compression performed
by WILLIAM (yet unpublished).

This concludes our presentation of incremental compression schemes
and their time complexities. We turn to an attempt to deepen our understanding
of what a feature is. Since features describe general non-random aspects
of a string, it turns out, there is a close relationship to the celebrated
Martin-Löf theory of randomness.

\section{Relationship to Martin-Löf randomness\label{subsec:Relationship-to-Martin-L=0000F6f}}

Recalling the definition of a feature, it comes to mind that it could
serve as a general algorithmic definition of an object's properties.
The expression $U\left(\left\langle r,f\right\rangle \right)=x$ means
that the feature is part of the description of object $x$. Properties
could be viewed as partial descriptions. The compression condition
demands that the property is not trivial in some sense. For example,
the partial description ``begins with $011010100$'' would violate
the compression condition, but it is not particularly interesting
and begs the question whether this partial description should be called
a property at all. After all, a property should be something that
demarcates a particular class of objects from all other objects in
a non-trivial way. In other words, an object possessing a property
should be rare in some sense and therefore compressible.

This idea is closely tied to the idea of Martin-Löf randomness. A
string is called Martin-Löf random, if it passes all randomness tests.
Recall the definition of a uniform test for randomness \citep[Definition 2.4.1]{li2009introduction}
($d(\cdot)$ measures the cardinality of a set):

\begin{dfn}[Randomness test]A total function $\delta:\mathcal{N}\rightarrow\mathcal{N}$
is a \textbf{uniform Martin-Löf test for randomness} if $\delta$
is lower semicomputable and $d\left(\left\{ x:l(x)=n,\;\delta(x)\ge m\right\} \right)\le2^{n-m}$,
for all $n$.\end{dfn}

If $\delta$ measures some non-random aspect of $x$, for example
the number of initial zeros, then the fraction of random strings with
high values of $\delta(x)$ should be low. Otherwise, the string is
unlikely to be random. In the present paper, the task of a feature
is to map out some non-random aspect of $x$. Therefore, there should
be some relationship between features and randomness tests, substantiated
by the following theorems:

\begin{thm}[From features to randomness tests]\label{thm:from-f-to-delta}For
each feature $f$ of some finite string, the function
\begin{equation}
\delta(x):=\begin{cases}
\max\left\{ l(x)-l(r)-1:f(r)=x,l(f)+l(r)<l(x)\right\}  & \mbox{if such an \ensuremath{r} exists}\\
0 & \mbox{otherwise}
\end{cases}\label{eq:ftest}
\end{equation}
is a uniform Martin-Löf test for randomness.\end{thm}\begin{pf}Fix
$f$. We define $\phi(t,x)$ as follows: For each $x$, run feature
$f$ for $t$ steps on each residual $r$ of length less than $l(x)-l(f)$.
If for any such input $r$ the computation halts with output $x$,
then define $\phi(t,x):=l(x)-l(r)-1$ using the shortest such $r$,
otherwise set $\phi(t,x):=0$. Clearly, $\phi(t,x)$ is recursive,
total, and monotonically nondecreasing with $t$ (for all $x$, $\phi(t',x)\ge\phi(t,x)$
if $t'>t$). The limit exists, since for each $x$ either no such
$r$ is found, making $\phi(t,x)=0$ for all $t$, or a shortest $r$
is found eventually. Therefore, $\lim_{t\rightarrow\infty}\phi(t,x)=\delta(x)$
and we have shown that $\delta$ is lower semicomputable.

Consider all $x$ with length $n$. The case $m=0$ is trivial, so
consider the case $m\geq1$. For each $x$ that meets condition $\delta(x)\ge m$
there has to exist some $r$ with $f(r)=x$, $l(f)+l(r)<l(x)$, and
$n-l(r)-1\ge m$. Therefore, $l(r)\leq n-m-1$ and the number of such
$r$ is bounded by $\sum_{i=0}^{n-m-1}2^{i}=2^{n-m}-1$. Since different
$x$'s require different $r$'s (they are executed on the same $f$),
the number of such $x$ is bounded by the same expression.\qed\end{pf}

\paragraph{Example 1}

Let $x=0^{a}w$. Clearly, there is a feature $f$ that takes residual
$r=\bar{a}w$ and computes $x$, if $a$ is big enough to fulfill
the compression condition. Since $l(r)=2l(a)+1+l(w)$ and $l(x)=a+l(w)$,
we obtain $\delta(x)\ge a-2l(a)-2$. For example, we can simply set
$\delta(x)=a$, which is clearly a randomness test that counts the
number of leading zeros of $x$ \citep[Example 2.4.1]{li2009introduction}.

\paragraph{Example 2}

Consider a string that has $a$ $1$'s at odd positions, $x=1x_{2}1x_{4}1x_{6}1x_{8}\ldots z$
and continues with some arbitrary string $z$ eventually. There is
a feature $f$ that takes residual $r=\bar{a}x_{2}x_{4}\cdots x_{2a}z$
and computes $x$, if $a$ is big enough to fulfill the compression
condition. Since $l(r)=2l(a)+1+a+l(z)$ and $l(x)=2a+l(z)$ we obtain
$\delta(x)\ge a-2l(a)-2$ again. We can choose
\begin{equation}
\delta(x):=a=\max\left\{ i:x_{1}=x_{3}=\cdots=x_{2i-1}=1\right\} \label{eq:ex1test}
\end{equation}
which is a randomness test \citep[Example 2.4.3]{li2009introduction}.

Conversely, there is also a map in the reverse direction, from randomness
tests to features:

\begin{thm}[From randomness tests to features]\label{thm:from-delta-to-f}For
each uniform and unbounded Martin-Löf test for randomness $\delta$,
there is a feature $f$ such that it is a feature of all $x$ fulfilling
$\delta(x)>l(f)$.\end{thm}\begin{pf}Let the set $V_{m}^{n}$ be
defined as
\begin{equation}
V_{m}^{n}:=\left\{ x:\delta(x)\ge m,l(x)=n\right\} 
\end{equation}
The lower semicomputability of $\delta$ implies that $V_{m}^{n}$
is recursively enumerable. We have defined $V_{m}^{n}$ such that
for any $x$ fulfilling condition $\delta(x)\ge m$ we have $x\in V_{m}^{l(x)}$
and $d\left(V_{m}^{l(x)}\right)\le2^{l(x)-m}$ with $m$ to be fixed
later. If $V_{m}^{l(x)}$ is not empty then $l(x)-m\geq0$. Let $\delta=\delta_{y}$
in the standard enumeration $\delta_{1},\delta_{2},\ldots$ of tests.
Given $y$, $m$ and $l(x)$, we have an algorithm to enumerate all
elements of $V_{m}^{l(x)}$. Together with the index $j$ of $x$
in the enumeration order of $V_{m}^{l(x)}$, this suffices to find
$x$. We pad the standard binary representation of $j$ with nonsignificant
zeros to a string $r=00\ldots0j$ of length $l(x)-m$. This is possible
since $d\left(V_{m}^{l(x)}\right)\le2^{l(x)-m}$, thus any index of
the element can be encoded by $l(r)=l(x)-m$ bits. The purpose of
changing $j$ to $r$ is that now the length $l(x)$ can be deduced
from $l(r)$ and $m$. In particular, we can encode $y$ and $m$
into a string $f$ corresponding to a Turing machine that computes
$x$ from input $r$. This shows the existence of a string $r$ with
$f(r)=x$ for any $x$ with $\delta(x)\ge m$ (for fixed $m$).

The compression condition becomes $l(x)>l(f)+l(r)=l(f)+l(x)-m$, hence
we require $l(f)<m$. Since it takes at most $2\log m$ bits to encode
$m$ into $f$ this inequality can always be fulfilled for large enough
$m$. Moreover, it is possible to fulfill $l(f)=m-1$ by adding some
unnecessary bits to $f$. Let us fix some appropriate $m$ and $f$
such that $l(f)=m-1$. Therefore, $f$ is indeed a feature of all
$x$ with $\delta(x)\ge m$, which is equivalent to $\delta(x)>l\left(f\right)$.
\qed\end{pf}

Intuitively, all strings $x$ of fixed length that fulfill $\delta(x)\ge m$
for some test $\delta$ possess some non-random aspect, and there
are few of them by definition of $\delta$. Therefore, a feature can
be constructed to enumerate them.

\paragraph{Example 3}

Consider again Example 2, now departing from the particular test in
eq.\ (\ref{eq:ex1test}) encoded by number $y$ in the standard enumeration
of tests. There are $2^{l(x)-a}$ strings of length $l(x)$ with 1's
at the first $a$ odd positions. Clearly, if we encode $y$ and $a$
into a function $f$, it can print $x$ given the index $j$ of $x$
in that set.

This map to Martin-Löf randomness tests establishes an important point.
While both features and randomness tests are able to measure the \textit{amount}
of randomness deficiency, features also describe the \textit{content}
of randomness deficiency, i.e.\ of a regularity. In this sense, features
appear more informative than randomness tests and may deserve the
status of the algorithmic formalization of the meaning of ``property''
of computable objects.

\section{Discussion}

We have presented a theory of incremental compression of arbitrary
finite data. It applies to any compressible data $x$ and suggests
to decompose the compression endeavor into small independent pieces:
the features. The main result, illustrated in Fig.\ \ref{fig:theorems}e),
shows that ultimately the length of the obtained description of $x$
by features $f_{1},\ldots,f_{s}$ and last residual $r_{s}$ reaches
the optimal Kolmogorov complexity $K(x)$ to logarithmic precision,
if the shortest possible features $f_{i}$ are selected at each incremental
compression step. Additionally, we have introduced a computable version
of the theory -- the ALgorithm for Incremental ComprEssion (ALICE).

\subsection{Relationship to file and image compression algorithms}

Each compression algorithm can be viewed as a descriptive map $f'$
of constant $O(1)$ length, which compresses $x$ and outputs a compressed
string -- a residual description $r$. Then the initial string $x$
can be restored by decompressing the residual, $x=f(r)$. In most
cases, algorithms are one-sided and perform one technique: Huffman
coding, arithmetic coding, delta encoding, Lempel-Ziv compression.
There are also algorithms like bzip2 and DEFLATE, which combine previously
mentioned techniques incrementally and are more complex practical
examples of the theory described here. For example, the DEFLATE algorithm
represents the initial string as a composition $\left(f_{H}\circ f_{LZ}\right)(r)$,
where $f_{H}$ can be viewed as a Huffman coding feature and $f_{LZ}$
-- a Lempel-Ziv compression feature. Other compression algorithms
like fractal compression can also be related to the present formalism.
When a picture is decoded as a set of rules for building a fractal
resembling a picture after some iterations of a fractal drawing, this
process could be represented as the application of the same feature
several times: $x=\left(f_{1}\circ\ldots\circ f_{1}\right)(r)$, where
$f_{1}$ represents only one iteration.

Overall, usual compression algorithms can achieve some compression
for a vast class of strings and may be expected to emerge at the first
steps of compression because of their small length. Nevertheless,
due to the narrowness of those methods, the achieved compression rate
can be expected to be far from the golden standard set by Kolmogorov
complexity $K(x)$, whereas the theory proposed here is fairly general
(see Subsection \ref{subsec:Generality}) and it reaches $K(x)$ with
certain precision (Theorem \ref{thm:compression_scheme}).

\subsection{Generality\label{subsec:Generality}}

There is probably a price to be paid for the gained efficiency. The
theory is only applicable to data possessing any properties, i.e.\
features at all. As the relationship to Martin-Löf randomness has
demonstrated, properties -- i.e.\ non-random aspects -- guarantee
the existence of features. Even though it is hard to imagine compressible
data without any features at all, it might exist in abundance. Of
course, compressible strings always possess at least one feature --
the universal feature if the string is long enough (Lemma \ref{lem:lemma_universal_feature})
or the singleton feature engulfing all of the information about $x$
(Lemma \ref{lem:lemma_total_feature}). These degenerate features
constitute extreme cases and contain either nonsignificant or all
of the information about $x$. However, the interesting question is
how many nondegenerate cases exist, in which the information in $x$
is divided into several chunks of intermediate size, which is expected
to lead to the highest boost in efficiency, according to eq.\ (\ref{eq:eff_boost}).
Unfortunately, in order to estimate this number, we would need to
know how feature lengths are distributed. If strings with these nondegenerate,
medium-sized features constitute a genuine subset of all data, it
would render our theory non-universal. From a practical perspective
however, the universe we inhabit seems to be teeming with features,
wherever we look. In this sense, our theory may not be a theory of
universal compression, but a compression theory for our universe.

\subsection{Machine learning and compression}

Machine learning models generalize better when the number of degrees
of freedom of the employed models is small and the size of the data
set large. In concordance with the reviewed Solomonoff theory of universal
induction the generalization ability of a model is firmly tied to
the idea of data compression. In fact, it is not an exaggeration to
say that data compression is an essential property of machine learning
in general, sometimes disguised as the minimum description length
principle, bias-variance trade-off, various regularization techniques
and model selection criteria \citep{schwarz1978estimating,bishop2006pattern,akaike1978new}.
For example, deep belief networks (DBNs) \citep{hinton2006fast} consist
of stacked autoencoders. In fact our theory of incremental compression
can be viewed as an algorithmic generalization of deep belief networks,
maybe even of deep learning in general, in so far as to show that
compressing data in small incremental steps (such as neuronal layers)
is a reasonable thing to do. Any transformation $f$ from a description
$r$ to the data $x$ can be viewed as a feature as long as some compression
is achieved. In the context of machine learning, often $f$ is fixed
after learning and is required to represent as set of data sets $x_{1},\ldots,x_{n}$
with the respective descriptions $r_{1},\ldots,r_{n}$: $f(r_{i})=x_{i}$.
In that respect, as long as $n$ is large and $l(r_{i})<l(x_{i})$,
compression is achieved. Our theory predicts that compression (and
thereby the generalization properties), will be best if the model
$f$ and the description $r$ of data in that model (including noise)
do not carry mutual information. This can be achieved by picking the
simplest possible model $f^{*}$ achieving compression. In the context
of DBNs, $f$ is the one-layer neural network generating visual neuron
patterns $x$ from the hidden neuron patterns $r$. This observation
raises doubts about whether a one-layer network can cover a broad
enough set of features for arbitrary data and is not too biased toward
a narrow class of transformations.

\subsection{Searching for features in practice}

It is important to ask how features can be found in practice. If exhaustive
search is to be used, the search might turn out to be slow. However,
as we have seen, the shortest feature $f^{*}$ and descriptive map
$f^{\prime*}$ are bounded by a constant for well-compressible strings.
Therefore, even though $f^{\ast}$ and $f^{\prime*}$ are incomputable
in general, an adaptation of the Coding Theorem Method \citep{zenil2018decomposition}
could be used since the halting problem has known solutions for simple
Turing machines. This in an interesting perspective, since a practical
implementation of incremental compression would be possible to some
extent without the loss of theoretical guarantees.

A straightforward method to search for features is to use some parametrized
family of functions $f_{\textbf{w}}$ employing e.g.\ gradient descent
on \ $E(\textbf{w}):=\left(f_{\textbf{w}}(r)-x\right)^{2}$. It turns
out that in this case an even shorter feature $f$ can be found since
the parameter $\textbf{w}$ can be moved into the residual: $f(\textbf{w},r)=x$
since $l(f)<l(f_{\textbf{w}})$ and in practice probably even $l(f)\ll l(f_{\textbf{w}})$.
This observation undermines the attempt to look for shortest features
using a parametrized family of functions and emphasizes that exhaustive
search might not be a bad idea, since the shortest features appear
to be very short and it allows the discovery of very different functions.
It does appear to work in practice \citep{franzetal2018,franz2019agi}.

Nevertheless, the incomputability of shortest features raises the
question about the consequences of picking \textit{not} the shortest
feature. From Theorem \ref{thm:independence-general} we know that
if $f$ is not the shortest, we risk wasting $d:=l(f)-l(f^{*})$ bits
of description length on the data: $K(f)+K(r)\lesssim K(x)+d$. If
the model class $f_{\textbf{w}}$ from which we pick the models is
not appropriate for the data, then the shortest model from this class
might still be much longer than the shortest model from the Turing
complete class, $l(f_{\textbf{w}}^{*})\gg l(f^{*})$, and much description
length would be wasted. Or, such as in the case of a single-layered
neural network, the reconstruction $f(r)$ might be far from the actual
data $x$ risking substantial information loss.

Apart from that, $f$ not being shortest might spoil its incompressibility
(Theorem \ref{thm:feature_incompressibility}) which might require
compressing $f$ further in order to achieve the smallest total description.

Note also that we should strive for lossless compression since the
residual $r$ contains both model parameters and noise. The reconstruction
$f(r)=x$ consists of computing the model prediction and adding the
noise to retrieve the data $x$ exactly. The ``noise'' is part of
$r$ since it may, in general, contain information not captured by
the first model $f_{1}$, but which might in turn be captured by the
next model $f_{2}$. This is reminiscent of principal component analysis,
where at each step the principal component of the a residual description
$r_{i}$ is searched for.

Another consequence of not picking the shortest feature (i.e.\ the
simplest model) could be that the residual description will be somehow
messed up. In other words, it could contain noise unrelated to $x$.
Algorithmically expressed, it is a question about the value of $K(r\mid x)$.
We have encountered this term in Theorem \ref{thm:no_superfluous_information}
about superfluous information. Interestingly, this can not happen
to a severe extent, even if the feature is not the shortest. This
follows from the proof of the above-mentioned theorem where so-called
\textit{first} features of a fixed length are search for. If we fix
the length of the feature $l(f)=n$, then $r$ can be computed from
$x$. Therefore, $K(r\mid x)\le K(r\mid x,n)+K(n)+O(1)\le K(n)+O(1)=O\left(\log l(f)\right)$,
which is not much. For example, in the context of DBNs, if the first
layer consists of weights $w_{1},\ldots,w_{N}$, then $l(f)\approx\sum_{i=1}^{N}l\left(D(w_{i})\right)$
where $D$ is some way to describe floats to some precision. If the
number of data-description pairs $(x_{i},r_{i})$ is large compared
to that, the superfluous information in $r$ will be small. Thus,
while we could spoil some description length by overlap $d$, the
residual will not be spoiled much, meaning that by compressing the
residual further we will still be mostly describing the original data
$x$ and not irrelevant information. It might be confounded by information
in $f$, but since the superfluous information in the latter is also
bounded by $O\left(\log l(f)\right)$, it does not matter much.

Another concern might be that an unfortunate autoencoder $(f_{1},f_{1}')$
might transform $x$ into a description $r_{1}$ whose shortest feature
$f_{2}^{*}$ might be long and therefore difficult to find. After
all, there is no guarantee, that there always exists a decomposition
of a large set of short features, which would be optimal from the
point of view of search efficiency. However, as we have seen in Theorem
\ref{thm:thm_short_features}, a string will always have a feature
whose length is bounded by a constant, if it is well-compressible.
If it is not well-compressible, not much can be done anyway, since
we are dealing with an almost random string.

It all sounds like good news, if we merely care about the amount of
information but not computation time. For example, if an unfortunate,
highly non-linear transformation $f'$ distorts $x$ into $r$, the
amount of ``distortion'' in $r$ is not high only because we assume
that we can find the reverse, ``fixing'' operation $f$ by merely
providing its length $l(f)$ to an algorithm that searches through
all $2^{l(f)}$ such reverse transformations until it finds the one
that generates $x$ from the distorted $r$. In practice, $l(f)$
is not available and executing $2^{l(f)}$ functions in parallel might
simply be too computationally expensive. Therefore, even if the shortest
program computing $r$, with length $K(r)$, is not much longer than
$K(x)-l(f^{*})\approx K(r)-K(f^{*},r\mid x)$, we might end up spending
much time ``unwinding'' the superfluous information $K(f^{*},r\mid x)$
that has been inserted into $r$. It is hence advisable to use short
descriptive maps $f'$ that are not able to inject much superfluous
information into $r$, due to the relationship $l(f^{\prime\ast})=K(r\mid x)$
in Lemma \ref{lem:lemma_feature_length_complexity}. In practice,
for example, in the case of linear regression $f'$ might be the least
squares algorithm that finds the parameters of a linear function such
as to minimize the reconstruction error. Apart from the parameters,
the residual $r$ will mostly consist of the reconstruction error
which is not spoiled much by the transformation and can be compressed
further if it contains information not captured by the linear function.

\section{Conclusions}

As we have seen in the discussion, the presented theory lays a foundation
for a more efficient and fairly general compression algorithm, aiming
to be applied in practice. Most importantly, since many machine learning
algorithms rely on good data compression the generality of the proposed
compression scheme could help those algorithms to overcome their narrowness
and improve their performance. In the context of the theory of universal
intelligence \citep{Hutter04uaibook} this theory could be a fruitful
way to derive more efficient formulations of the generally intelligent
AIXI agent.

\section*{Acknowledgments}

We are grateful to Kenshi Miyabe for the comment expressed in Remark
\ref{rem:remark2}. We thank Valerie Barkova for making the figures.
\textbf{Author contributions}: A.F. had the original idea of incremental
compression and developed the main theory and Greedy-ALICE. He also
had the initial idea about the relationship to Martin-Löf randomness
and wrote most of the paper. O.A. corrected the proofs for Theorems
\ref{thm:feature_incompressibility} and \ref{thm:no_superfluous_information},
developed the $b-$compressible theory (Subsections \ref{subsec:On-the-number}-\ref{subsec:Comparison-of-compression})
and the main part of the Theorems \ref{thm:from-f-to-delta} and \ref{thm:from-delta-to-f}
on randomness tests. R.S. found Theorem \ref{thm:number_of_features},
helped out on the orthogonality Theorem \ref{thm:orthogonality} and
has developed ALICE and the respective lemma and theorem. \textbf{Funding}:
This work was privately funded by Arthur Franz and Michael Löffler
within the OCCAM research laboratory.

\bibliographystyle{plainnat}
\bibliography{incrementalcompression}

\begin{thebibliography}{30}
\providecommand{\natexlab}[1]{#1}
\providecommand{\url}[1]{\texttt{#1}}
\expandafter\ifx\csname urlstyle\endcsname\relax
  \providecommand{\doi}[1]{doi: #1}\else
  \providecommand{\doi}{doi: \begingroup \urlstyle{rm}\Url}\fi

\bibitem[Akaike(1978)]{akaike1978new}
Hirotugu Akaike.
\newblock A new look at the statistical model identification.
\newblock \emph{Automatica}, 19\penalty0 (6):\penalty0 465--471, 1978.

\bibitem[Barlow(1961)]{barlow1961possible}
Horace~B. Barlow.
\newblock Possible principles underlying the transformation of sensory
  messages.
\newblock \emph{Sensory communication}, 1:\penalty0 217--234, 1961.

\bibitem[Barmpalias and Lewis-Pye(2019)]{barmpalias2019compression}
George Barmpalias and Andrew Lewis-Pye.
\newblock Compression of data streams down to their information content.
\newblock \emph{IEEE Transactions on Information Theory}, 65\penalty0
  (7):\penalty0 4471--4485, 2019.

\bibitem[Bishop(2006)]{bishop2006pattern}
Christopher~M. Bishop.
\newblock \emph{Pattern recognition and machine learning}.
\newblock Springer, 2006.

\bibitem[Franz(2016)]{franz2016some}
Arthur Franz.
\newblock Some theorems on incremental compression.
\newblock In \emph{International Conference on Artificial General
  Intelligence}, pages 74--83. Springer, 2016.

\bibitem[Franz et~al.(2018)Franz, L{\"o}ffler, Antonenko, Gogulya, and
  Zaslavskyi]{franzetal2018}
Arthur Franz, Michael L{\"o}ffler, Alexander Antonenko, Victoria Gogulya, and
  Dmytro Zaslavskyi.
\newblock Introducing {WILLIAM}: a system for inductive inference based on the
  theory of incremental compression.
\newblock In \emph{International Conference on Computer Algebra and Information
  Technology}, 2018.

\bibitem[Franz et~al.(2019)Franz, Gogulya, and L{\"o}ffler]{franz2019agi}
Arthur Franz, Victoria Gogulya, and Michael L{\"o}ffler.
\newblock {WILLIAM}: A monolithic approach to {AGI}.
\newblock In \emph{International Conference on Artificial General
  Intelligence}, pages 44--58. Springer, 2019.

\bibitem[Hinton and Salakhutdinov(2006)]{hinton2006reducing}
Geoffrey~E. Hinton and Ruslan~R. Salakhutdinov.
\newblock Reducing the dimensionality of data with neural networks.
\newblock \emph{Science}, 313\penalty0 (5786):\penalty0 504--507, 2006.

\bibitem[Hinton et~al.(2006)Hinton, Osindero, and Teh]{hinton2006fast}
Geoffrey~E. Hinton, Simon Osindero, and Yee-Whye Teh.
\newblock A fast learning algorithm for deep belief nets.
\newblock \emph{Neural computation}, 18\penalty0 (7):\penalty0 1527--1554,
  2006.

\bibitem[Hutter(2002)]{Hutter:01fast}
Marcus Hutter.
\newblock The fastest and shortest algorithm for all well-defined problems.
\newblock \emph{International Journal of Foundations of Computer Science},
  13\penalty0 (3):\penalty0 431--443, June 2002.
\newblock URL \url{http://www.hutter1.net/ai/pfastprg.htm}.

\bibitem[Hutter(2005)]{Hutter04uaibook}
Marcus Hutter.
\newblock \emph{Universal Artificial Intelligence: Sequential Decisions based
  on Algorithmic Probability}.
\newblock Springer, Berlin, 2005.
\newblock ISBN 3-540-22139-5.
\newblock \doi{10.1007/b138233}.
\newblock 300 pages, http://www.hutter1.net/ai/uaibook.htm.

\bibitem[Jr.(2003)]{gauch2003scientific}
Hugh G.~Gauch Jr.
\newblock \emph{Scientific method in practice}.
\newblock Cambridge University Press, 2003.

\bibitem[Kemp and Regier(2012)]{kemp2012kinship}
Charles Kemp and Terry Regier.
\newblock Kinship categories across languages reflect general communicative
  principles.
\newblock \emph{Science}, 336\penalty0 (6084):\penalty0 1049--1054, 2012.

\bibitem[Knill and Richards(1996)]{knill1996perception}
David~C. Knill and Whitman Richards.
\newblock \emph{Perception as Bayesian inference}.
\newblock Cambridge University Press, 1996.

\bibitem[Kolmogorov(1965)]{kolmogorov1965three}
Andrei~N. Kolmogorov.
\newblock Three approaches to the definition of the concept "quantity of
  information".
\newblock \emph{Problemy peredachi informatsii}, 1\penalty0 (1):\penalty0
  3--11, 1965.

\bibitem[Legg and Hutter(2007)]{legg2007universal}
Shane Legg and Marcus Hutter.
\newblock Universal intelligence: A definition of machine intelligence.
\newblock \emph{Minds and machines}, 17\penalty0 (4):\penalty0 391--444, 2007.

\bibitem[Levin(1973)]{levin1973universal}
Leonid~A Levin.
\newblock Universal sequential search problems.
\newblock \emph{Problemy Peredachi Informatsii}, 9\penalty0 (3):\penalty0
  115--116, 1973.

\bibitem[Li and Vit{\'a}nyi(2009)]{li2009introduction}
Ming Li and Paul~MB Vit{\'a}nyi.
\newblock \emph{An introduction to Kolmogorov complexity and its applications}.
\newblock Springer, 2009.

\bibitem[MacKay(2003)]{mackay2003information}
David J.~C. MacKay.
\newblock \emph{Information theory, inference and learning algorithms}.
\newblock Cambridge university press, 2003.

\bibitem[Olshausen and Field(1996)]{olshausen1996emergence}
Bruno~A. Olshausen and David~J. Field.
\newblock Emergence of simple-cell receptive field properties by learning a
  sparse code for natural images.
\newblock \emph{Nature}, 381\penalty0 (6583):\penalty0 607, 1996.

\bibitem[Potapov and Rodionov(2014)]{potapovSSsearch}
Alexey Potapov and Sergey Rodionov.
\newblock Making universal induction efficient by specialization.
\newblock In \emph{International Conference on Artificial General
  Intelligence}, pages 133--142. Springer, 2014.

\bibitem[Rissanen(1978)]{rissanen1978modeling}
Jorma Rissanen.
\newblock Modeling by shortest data description.
\newblock \emph{Automatica}, 14\penalty0 (5):\penalty0 465--471, 1978.

\bibitem[Schmidhuber(2004)]{schmidhuber2004optimal}
J{\"u}rgen Schmidhuber.
\newblock Optimal ordered problem solver.
\newblock \emph{Machine Learning}, 54\penalty0 (3):\penalty0 211--254, 2004.

\bibitem[Schmidhuber et~al.(1997)Schmidhuber, Zhao, and
  Wiering]{schmidhuber1997shifting}
J{\"u}rgen Schmidhuber, Jieyu Zhao, and Marco Wiering.
\newblock Shifting inductive bias with success-story algorithm, adaptive levin
  search, and incremental self-improvement.
\newblock \emph{Machine Learning}, 28\penalty0 (1):\penalty0 105--130, 1997.

\bibitem[Schwarz(1978)]{schwarz1978estimating}
Gideon Schwarz.
\newblock Estimating the dimension of a model.
\newblock \emph{The {A}nnals of {S}tatistics}, 6\penalty0 (2):\penalty0
  461--464, 1978.

\bibitem[Shen et~al.(2017)Shen, Uspensky, and Vereshchagin]{shen2017kolmogorov}
Alexander Shen, Vladimir~A. Uspensky, and Nikolay Vereshchagin.
\newblock \emph{Kolmogorov complexity and algorithmic randomness}, volume 220.
\newblock American Mathematical Society, 2017.

\bibitem[Solomonoff(1964{\natexlab{a}})]{solomonoff1964formal1}
Ray~J. Solomonoff.
\newblock A formal theory of inductive inference. {P}art {I}.
\newblock \emph{Information and control}, 7\penalty0 (1):\penalty0 1--22,
  1964{\natexlab{a}}.

\bibitem[Solomonoff(1964{\natexlab{b}})]{solomonoff1964formal2}
Ray~J. Solomonoff.
\newblock A formal theory of inductive inference. {P}art {II}.
\newblock \emph{Information and control}, 7\penalty0 (2):\penalty0 224--254,
  1964{\natexlab{b}}.

\bibitem[Solomonoff(1978)]{solomonoff1978complexity}
Ray~J. Solomonoff.
\newblock Complexity-based induction systems: comparisons and convergence
  theorems.
\newblock \emph{IEEE transactions on Information Theory}, 24\penalty0
  (4):\penalty0 422--432, 1978.

\bibitem[Zenil et~al.(2018)Zenil, Hern{\'a}ndez-Orozco, Kiani, Soler-Toscano,
  Rueda-Toicen, and Tegn{\'e}r]{zenil2018decomposition}
Hector Zenil, Santiago Hern{\'a}ndez-Orozco, Narsis~A Kiani, Fernando
  Soler-Toscano, Antonio Rueda-Toicen, and Jesper Tegn{\'e}r.
\newblock A decomposition method for global evaluation of {S}hannon entropy and
  local estimations of algorithmic complexity.
\newblock \emph{Entropy}, 20\penalty0 (8):\penalty0 605, 2018.

\end{thebibliography}

\end{document}